\documentclass[11pt]{article}
\usepackage{amsmath,amssymb}
\makeatletter\@addtoreset {equation}{section}\makeatother

\newtheorem{theo}{Theorem}
\newtheorem{lem}{Lemma}[section]

\newtheorem{prop}{Proposition}[section]

\newtheorem{rem}{Remark}[section]
\newenvironment{Proof}
{\begin{trivlist} \item[]{\bf Proof. }}%
{\hspace*{\fill}$\rule{.3\baselineskip}{.35\baselineskip}$\end{trivlist}}

\newcommand{\R}{\mathbb{R}}

\newcommand{\N}{\mathbb{N}}

\renewcommand{\geq}{\geqslant}
\renewcommand{\leq}{\leqslant}

\renewcommand{\phi}{\varphi}
\newcommand{\be}{\begin{eqnarray}}
\newcommand{\ee}{\end{eqnarray}}

\newcommand{\eps}{\varepsilon}

\newcommand{\rqe}{{R}^\eps}
\newcommand{\fqe}{{Q}^\eps}
\newcommand{\met}{\tilde{M}^\eps}
\newcommand{\qet}{{Q}^\eps}
\newcommand{\mne}{\tilde{\mu}_n^\eps}

\begin{document}

\title{\bf On the Thomas--Fermi ground state \\ in a harmonic potential}

\author{Cl\'ement Gallo and Dmitry Pelinovsky \\
{\small Department of Mathematics, McMaster
University, Hamilton, Ontario, Canada, L8S 4K1}  }

\date{\today}
\maketitle

\begin{abstract}
We study nonlinear ground states of the Gross--Pitaevskii
equation in the space of one, two and three dimensions with a radially
symmetric harmonic potential. The Thomas--Fermi
approximation of ground states on various spatial scales was
recently justified using variational methods. We justify here
the Thomas--Fermi approximation on an uniform spatial scale using
the Painlev\'{e}-II equation. In the space of one dimension, these
results allow us to characterize the distribution of eigenvalues in
the point spectrum of the Schr\"{o}dinger operator associated with the
nonlinear ground state. 

\end{abstract}

\section{Introduction}

Recent experiments with Bose--Einstein condensates \cite{PitStr}
have stimulated new interest in the Gross--Pitaevskii equation
with a harmonic potential. We take this equation in the form
\begin{equation}
\label{GP} i u_t + \eps^2 \Delta u + (1 - |x|^2) u - |u|^2 u = 0,
\quad x \in \R^d, \quad t \in \R_+,
\end{equation}
where the space dimension is $d$ is one, two or three, $u(x,t) \in \mathbb{C}$ is the wave function of the repulsive
Bose gas in the mean-field approximation, and $\eps$ is a small
parameter that corresponds to the Thomas--Fermi approximation of a
nearly compact atomic cloud \cite{Fermi,Thomas}.

A ground state of the Bose-Einstein condensate is a positive,
time-independent solution $u(x,t) = \eta_{\eps}(x)$ of the Gross--Pitaevskii equation (\ref{GP}).
More precisely, $\eta_{\eps}: \mathbb{R}^d \mapsto \mathbb{R}$ satisfies the stationary
Gross--Pitaevskii equation
\begin{equation}
\label{stationaryGP}
\eps^2\Delta\eta_\eps(x)+(1-|x|^2)\eta_\eps(x)-\eta_\eps^3(x)=0,  \quad
x \in \R^d,
\end{equation}
$\eta_\eps(x) > 0$ for all $x \in \R^d$, and $\eta_\eps$ has a finite
energy $E_\eps(\eta_\eps)$, where $E_\eps$ is given by
$$E_\eps(u)=\int_{\R^d}\left(\eps^2|\nabla u|^2+(|x|^2-1)u^2+\frac{1}{2}u^4\right)dx.$$

For $d=2$, existence and
uniqueness of a radially symmetric ground state $\eta_{\eps}$ for a fixed, sufficiently   small $\eps
> 0$ is proven in Theorem 2.1 of Ignat \& Millot \cite{IM}
similarly to earlier works of Brezis \& Oswald \cite{BO} and
Aftalion, Alama, \& Bronsard \cite{AAB} in bounded domains. It is also shown in \cite{IM}
that $\eta_{\eps}(x)$ converges to $\eta_0(x)$ as $\eps \to 0$ for all
$x \in \R^2$, where $\eta_0$ is the Thomas--Fermi's compact function
\begin{equation}
\label{Thomas-Fermi}
\eta_0(x) = \left\{ \begin{array}{cl} (1 - |x|^2)^{1/2} & \text{for }  |x| < 1, \\
0 & \text{for }  |x| > 1. \end{array} \right.
\end{equation}
To be precise, Proposition 2.1 of \cite{IM} states that for $d=2$, $\eps>0$
sufficiently small,
\begin{eqnarray}
0 \leq \eta_{\eps}(x) \leq C \eps^{1/3} \exp\left(\frac{1 - |x|^2}{4
\eps^{2/3}}\right) & \mbox{for} & |x| \geq 1,\label{IM1} \\
0 \leq (1-|x|^2)^{1/2} - \eta_{\eps}(x) \leq C \eps^{1/3} (1 - |x|^2)^{1/2}&
\mbox{\rm for} & |x| \leq 1 - \eps^{1/3},\label{IM2}
\end{eqnarray}
and
\begin{eqnarray}
\| \eta_{\eps} - \eta_0 \|_{C^1(K)} \leq C_K \eps^2,\label{IM3}
\end{eqnarray}
where $K$ is any compact
subset of $\{x\in\R^2 : |x|<1\}$ and $C$ and $C_K$ are
$\eps$-independent positive constants.  
The method used by Ignat \& Millot in the case $d=2$ to prove the existence of
a radially symmetric ground state $\eta_\eps$ can be extended to the
cases $d=1,3$, even 
though the uniqueness of the ground state does not follow from
\cite{IM} for $d=3$. We are concerned here with a uniform asymptotic
approximation of the ground state $\eta_{\eps}$ on $\R^d$, in the
limit $\eps\to 0$, for $d=1,2,3$. 

At least two attempts have been
made in physics literature \cite{BTNN,KK} to establish connection
between the nonlinear ground state $\eta_{\eps}$ for $d=1$ and solutions of
the Painlev\'e-II equation
\begin{equation}
\label{Peinleve} 4 \nu''(y) + y \nu(y) - \nu^3(y) = 0,\quad
y\in\R.
\end{equation}
This equation arises as the formal limit as $\eps \to 0$ of the
differential equation satisfied by $\nu_\eps$:
$$
4 (1 - \eps^{2/3} y) \nu_{\eps}''(y) - 2 \eps^{2/3} d\nu_{\eps}'(y)
+ y \nu_{\eps}(y) - \nu_{\eps}^3(y) = 0, \quad
y \in (-\infty,\eps^{-2/3}),
$$
where $\nu_\eps$ is defined by
\begin{equation}
\label{asymptotic-transformation}
\eta_\eps(x)=\eps^{1/3}\nu_\eps(y), \quad y =
\frac{1-|x|^2}{\eps^{2/3}}.
\end{equation}
The convergence of $\eta_{\eps}$ to $\eta_0$ as $\eps \to 0$
suggests that we should consider the Hasting--McLeod
solution $\nu_0$ of the Painlev\'e-II equation \cite{HM}, which is
the unique solution of (\ref{Peinleve}) such that
$$
\nu_0(y) \sim
y^{1/2} \quad \mbox{as} \quad y\to +\infty \quad \mbox{and} \quad
\nu_0(y)\to 0 \quad \mbox{as} \quad y\to-\infty.
$$
In both papers \cite{BTNN,KK}, the asymptotic solution $\eta_{\eps}$
is constructed at three spatial scales
$$
{\rm I} : \; |x| \leq 1 - \eps^{2/3}, \quad {\rm II} : \;|x| \in
(1-\eps^{2/3},1+\eps^{2/3}), \quad \mbox{and} \quad {\rm III} : \;
|x| \geq 1 + \eps^{2/3}.
$$
Solutions of the Painlev\'e-II equation (\ref{Peinleve}) are used
at the intermediate scale II for matching conditions and
connection formulas between the WKB solutions at the inner
scale I and the Airy function solutions at the outer scale III.
The same formal approach is also developed in
\cite{ZAKP} for approximations of
excited states of the stationary Gross--Pitaevskii equation in the
case $d=1$.

We address the problem of uniform asymptotic approximations of the
ground state $\eta_{\eps}$ of the stationary
Gross--Pitaevskii equation (\ref{stationaryGP}) using the
Hasting--McLeod solution of the Painlev\'e-II equation
(\ref{Peinleve}). Our main result (Theorem \ref{theorem-main}) in
Section 2 establishes this approximation on a rigorous level. In the case when $d=1$, we
also study eigenvalues of the Schr\"{o}dinger operator
$$
L_+^{\eps} = -\eps^2 \partial_x^2 + V_{\eps}(x), \quad V_{\eps}(x)= 3 \eta^2_{\eps}(x) - 1 + x^2,
$$
that arises in the linearization of the stationary Gross--Pitaevskii equation (\ref{stationaryGP})
at the ground state $\eta_{\eps}$. We prove in Section 3 that the
spectrum of $L_+^{\eps}$ in $L^2(\R)$ consists of an infinite sequence of positive eigenvalues
$\{ \lambda_n^\eps \}_{n \geq 1}$ such that for any fixed integer $k \geq 1$,
\begin{equation}
\label{asympt-eigenvalues}
\lambda_{2k-1}^\eps, \lambda_{2k}^\eps \sim \mu_k \eps^{2/3} \quad \mbox{\rm as} \quad \eps \to 0,
\end{equation}
where $\mu_k$ is the $k^{\text{th}}$ eigenvalue of the Schr\"{o}dinger operator
$$
M_0 = -4 \partial_y^2 + W_0(y), \quad W_0(y) = 3 \nu_0(y) - y.
$$
We note that $M_0$ arises in the linearization of the
Painlev\'{e}-II equation (\ref{Peinleve}) at the Hasting--McLeod
solution $\nu_0$. Therefore, the scaling transformation
(\ref{asymptotic-transformation}) leading to the Painlev\'{e}-II
equation (\ref{Peinleve}) becomes useful for analysis of
eigenvalues of the Schr\"{o}dinger operator $L^{\eps}_+$.

It is clear from the shape of $\eta_{\eps}$ that the operator
$L^{\eps}_+$ has a double-well potential $V_{\eps}(x)$ with two
symmetric minima converging to $\pm 1$ as $\eps\to 0$, while the operator
$M_0$ has a single-well potential $W_0(y)$. These facts explain both the asymptotic
correspondence between eigenvalues of $L_+^{\eps}$ and $M_0$ and
the double degeneracy of each pair of eigenvalues in the
asymptotic limit (\ref{asympt-eigenvalues}). Formal results of the
semi-classical theory for the operator $L_+^{\eps}$ are collected
in Section 4.

While a different technique is exploited in our previous work
\cite{GalPel}, the result (\ref{asympt-eigenvalues}) provides the
same kind of asymptotic behaviour for the smallest eigenvalue of
$L_+^\eps$ as the one we obtained for the lowest
eigenvalue of the simplified operator
$$
\tilde{L}_+^{\eps} = -\eps^2 \partial_x^2 + V_0(x), \quad V_0(x)
= 3 \eta^2_0(x) - 1 + x^2.
$$

The spectral stability of the ground state in the Gross--Pitaevskii equation
(\ref{GP}) is deducted from the analysis of the symplectically coupled eigenvalue problem for
Schr\"{o}dinger operators $L_+^{\eps}$ and $L^{\eps}_-$, where
$$
L_-^{\eps} = -\eps^2 \partial_x^2 + \tilde{V}_{\eps}(x), \quad
\tilde{V}_{\eps}(x) = \eta^2_{\eps}(x) - 1 + x^2 = \frac{\eps^2
\eta_{\eps}''(x)}{\eta_{\eps}(x)}.
$$
Unfortunately, the asymptotic scaling
(\ref{asymptotic-transformation}) leading to the Painlev\'{e}-II
equation (\ref{Peinleve}) does not give a correct scaling of the
eigenvalues of $L_-^{\eps}$ nor the eigenvalues of the spectral
stability problem because the potential
$\tilde{V}_{\eps}(x)$ is a single well with a nearly flat bottom on
the interval $[-1,1]$, which is
mapped to
$[0,\eps^{-2/3}]$ by the change of variable $y=(1-x^2)/\eps^{2/3}$. Analysis of the eigenvalues of the spectral 
stability problem and construction of excited states of the
stationary Gross--Pitaevskii equation are two open problems beyond
the scope of this article.

\paragraph{Notations.}
If $A$ and $B$ are two quantities depending on a parameter $\eps$
belonging to a neighborhood $\mathcal{E}$ of 0, 
\begin{itemize}
\item $A(\eps) \lesssim B(\eps)$ indicates that there exists a positive
constant $C$ such that
$$
A(\eps) \leq C B(\eps) \quad \mbox{for every} \; \eps\in \mathcal{E}.
$$
\item $A(\eps)\underset{\eps\to 0}{\sim}B(\eps)$ if
$A(\eps)/B(\eps)\to 1$ as $\eps\to 0$
\item $A(\eps)=\mathcal{O}(B(\eps))$ as $\eps\to 0$ if
$A(\eps)/B(\eps)$ remains bounded as $\eps\to 0$.
\end{itemize}

Let $F(x)$ be a function defined in a neighborhood of $\infty$.
Given $\alpha \in \R$, $\{ f_m \}_{m \in \N} \in \R$, and $\gamma > 0$, the
notation
\be\label{defasympseries}
F(x) \underset{x \to \infty}{\approx} x^\alpha\sum_{m=0}^\infty f_m x^{-\gamma m}\nonumber
\ee
means that for every $M\in\N$,
$$
F(x)-x^\alpha\sum_{m=0}^M f_m x^{-\gamma
  m}=\mathcal{O}(x^{\alpha-\gamma(M+1)})\quad \text{as }x\to\infty,
$$
and, moreover, that the asymptotic series can be differentiated term by term.

We use the following spaces:
\begin{itemize}
\item $H^\infty(\R)=\underset{s\geq 0}{\cap}H^s(\R)$, where
$H^s(\R)$ is the standard Sobolev space. 
\item $L^2_r(\R^d)$ is the subspace of radially symmetric functions in
$L^2(\R^d)$. Note that if $f(|\cdot|)\in L^2_r(\R^d)$, then
$$\|f(|\cdot|)\|_{L^2(\R^d)}=|\mathbb{S}^{d-1}|\int_0^\infty
r^{d-1}|f(r)|^2dr,$$
where $|\mathbb{S}^{d-1}|$ is the surface of the unit sphere in
$\R^d$. Similarly, $|\mathbb{B}^d|$ is the volume of the unit ball in $\R^d$.
\end{itemize}

\section{Uniform asymptotic expansion of $\eta_{\eps}$}

In what follows, $d=1,2$ or $3$ and $\eps>0$ is sufficiently small such that, as it is
proved in Theorem 2.1 of \cite{IM}, there exists a positive classical
solution $\eta_\eps$ of
\begin{equation}
\label{eta-eq}
\eps^2\Delta\eta_\eps(x)+(1-|x|^2)\eta_\eps(x)-\eta_\eps^3(x)=0,\quad
x\in\R^d.
\end{equation}
Moreover, this ground state $\eta_\eps$ is radially symmetric, so that we can define a function $\nu_\eps$ on
$J_{\eps} := (-\infty,\eps^{-2/3}]$ by
\begin{equation}
\label{asymptotic-scaling}
\eta_\eps(x)=\eps^{1/3}\nu_\eps\left(\frac{1-|x|^2}{\eps^{2/3}}\right),
\quad x\in\R^d.
\end{equation}
Let $y = (1-|x|^2)/\eps^{2/3}$ be a new variable. Notice
that $y$ covers once $J_{\eps}$ as $|x|$ covers
$\R_+$. It is equivalent for
$\eta_\eps$ to solve (\ref{eta-eq}) and for $\nu_{\eps}$ to solve
the differential equation
\begin{equation}
\label{nu-eq}
4 (1 - \eps^{2/3} y) \nu_{\eps}''(y) - 2 \eps^{2/3}d \nu_{\eps}'(y) + y \nu_{\eps}(y) - \nu_{\eps}^3(y) = 0, \quad
y \in J_{\eps}.
\end{equation}
Let $N\geq 0$ be an integer. We look for $\nu_\eps$ using the form
\be\label{exp}
\nu_\eps(y)=\sum_{n=0}^N\eps^{2n/3}\nu_n(y)+\eps^{2(N+1)/3}R_{N,\eps}(y),
\quad y \in J_{\eps}.
\ee
Expansion (\ref{exp}) provides a solution of equation (\ref{nu-eq}) if $\{\nu_n\}_{0\leq
  n\leq N}$ and $R_{N,\eps}$ satisfy equations (\ref{P2-eq}), (\ref{nun-eq}) and
(\ref{r-eq}) below.
\begin{itemize}
\item $\nu_0$ solves the Painlev\'e-II equation
\begin{equation}
\label{P2-eq}
4 \nu_0''(y) + y \nu_0(y) - \nu^3_0(y) = 0, \quad y \in \mathbb{R},
\end{equation}
\item for $1\leq n\leq N$, $\nu_n$ solves
\be\label{nun-eq}
-4\nu_n''(y) + W_0(y)\nu_n(y) =F_n(y),\quad y\in \R,
\ee
where
$$
W_0(y) = 3 \nu_0^2(y) - y
$$
and
$$
F_n(y)=-\!\!\!\!\!\!\!\!\underset{\tiny{\begin{array}{c}n_1,n_2,n_3<n\\n_1+n_2+n_3=n\end{array}}}{\sum}\!\!\!\!\nu_{n_1}(y)\nu_{n_2}(y)\nu_{n_3}(y)-2d\nu_{n-1}'(y)-4y\nu_{n-1}''(y),$$
\item $R_{N,\eps}$ solves \be \label{r-eq}
-4(1-\eps^{2/3}y)R_{N,\eps}''+2\eps^{2/3}dR_{N,\eps}' + W_0 R_{N,\eps}=F_{N,\eps}(y,R_{N,\eps}),
\quad y \in J_{\eps},
\ee
where \be
\lefteqn{F_{N,\eps}(y,R)\ =\
-(4y\nu_N''+2d\nu_N')-\sum_{n=0}^{2N-1}\eps^{2n/3}\underset{\tiny{\begin{array}{c}n_1+n_2+n_3=n+N+1\\0\leq
      n_1,n_2,n_3\leq
      N\end{array}}}{\sum}\nu_{n_1}\nu_{n_2}\nu_{n_3}}\nonumber\\
&&-\left(3\sum_{n=1}^{2N}\eps^{2n/3}\underset{\tiny{\begin{array}{c}n_1+n_2=n\\0\leq
      n_1,n_2\leq N\end{array}}}{\sum}\nu_{n_1}\nu_{n_2}\right)R-
      \left(3\sum_{n=N+1}^{2N+1}\eps^{2n/3}\nu_{n-(N+1)}\right)R^2-\eps^{4(N+1)/3}R^3.\nonumber
\ee
\end{itemize}
Notice that for $0\leq n\leq N$, $\nu_n(y)$ is defined for all $y\in
\R$ and does not depend
on $\eps$, whereas $R_{N,\eps}(y)$ is a priori
only defined for $y\in J_{\eps}$.

Appropriate solutions of system (\ref{P2-eq}), (\ref{nun-eq}) and
(\ref{r-eq}) enable us to prove the following theorem.

\begin{theo}
\label{theorem-main}
Let $\nu_0$ be the unique solution of the Painlev\'e II equation
(\ref{P2-eq}) such that
$$
\nu_0(y) \sim y^{1/2} \quad \mbox{as} \quad y\to +\infty \quad \mbox{and} \quad
\nu_0(y)\to 0 \quad \mbox{as} \quad y\to-\infty.
$$
For $n\geq 1$, there exists a unique solution $\nu_n$ of equation
(\ref{nun-eq}) in $H^\infty(\R)$. For every $N\geq
0$, there exists $\eps_N > 0$ and $C_N>0$ such that for every
$0<\eps<\eps_N$, there is a solution $R_{N,\eps}\in
\mathcal{C}^\infty\cap L^\infty(J_{\eps})$ of equation (\ref{r-eq}) with
$$
\|R_{N,\eps}\|_{L^\infty(J_{\eps})} \leq C_N\eps^{-(d-1)/3} \quad
\text{and}\quad
x\mapsto R_{N,\eps}
\left(\frac{1-|x|^2}{\eps^{2/3}}\right)\in H^2(\R^d),
$$
such that
\be\label{expetaeps}
\eta_\eps(x)=\eps^{1/3}\sum_{n=0}^N\eps^{2n/3}\nu_n\left(
\frac{1-|x|^2}{\eps^{2/3}}\right)+\eps^{2N/3+1}R_{N,\eps}
\left(\frac{1-|x|^2}{\eps^{2/3}}\right),\quad x\in \R^d
\ee
is a ground state of equation (\ref{eta-eq}).
\end{theo}

\begin{rem} For $d=3$, the remainder term in (\ref{expetaeps})
may have the same order as the last term in the sum, because of the growth
of the upper bound on $\|R_{N,\eps}\|_{L^\infty(J_{\eps})}$ as
$\eps\downarrow 0$. 
\end{rem}
\begin{rem} For $d=1,2$, the ground state we find in
  Theorem \ref{theorem-main} is the unique ground state of equation
  (\ref{eta-eq}), thanks to the uniqueness result proved in
  \cite{IM}. For $d=3$, it is not clear whether the ground state of Theorem
  \ref{theorem-main} coincides with the one obtained by the
  method of Ignat and Millot in \cite{IM}, because uniqueness of a
  ground state does not follow from \cite{IM}.
\end{rem}
The proof of Theorem \ref{theorem-main} is described in the following three subsections. Notice first that it is sufficient to prove the Theorem for an arbitrarily large value of $N$. Indeed, for every integer $N_0>0$, the result of the Theorem for $N<N_0$ is a direct consequence of the result for $N=N_0$. Also, for convenience, we shall assume in the sequel that $N\geq 2$.

\subsection{Construction of $\nu_n$ for  $0\leq n\leq N$}

We are looking for a solution $\nu_{\eps}(y)$ of equation (\ref{nu-eq})
that satisfies the following limit as $\eps \to 0$:
$$\eps^{1/3}\nu_{\eps}(\eps^{-2/3}(1-x^2)) \underset{\eps\to
  0}{\longrightarrow} \left\{\begin{array}{cl}(1-x^2)^{1/2} &
    \text{for }x\in [-1,1],\\ 0 & \text{for } |x|\geq 1.\end{array}\right. $$
Therefore, we choose $\nu_0(y)$ to be the unique solution of the
Painlev\'e-II equation (\ref{P2-eq}) that satisfies the asymptotic
behavior $\nu_0(y) \sim y^{1/2}$ as $y\to +\infty$ and converges to
zero as $y\to-\infty$. Existence and uniqueness of this solution
are proved by Hastings \& McLeod \cite{HM}. Asymptotic behaviour
of $\nu_0(y)$ as $y\to\pm\infty$ is described in more details
in Theorem 11.7 of \cite{FIKN}. These results are
combined together in the following proposition.

\begin{prop}\cite{HM,FIKN}
The Painlev\'e-II equation
$$
4 \nu''(y) + y \nu(y) - \nu^3(y) = 0, \quad y \in \mathbb{R},
$$
admits a unique solution $\nu_0\in \mathcal{C}^{\infty}(\R)$ such
that
$$
\nu_0(y) \sim y^{1/2} \quad \mbox{as} \quad y\to +\infty \quad
\mbox{and} \quad \nu_0(y)\to 0 \quad \mbox{as} \quad y\to-\infty.
$$
Moreover, $\nu_0$ is strictly increasing on $\R$, $\nu_0''$ has
exactly one zero on $\R$, which is an inflection point of $\nu_0$.
The behaviour of $\nu_0$ as $y\to-\infty$ is described by
\be\label{asympnu0-}
\nu_0(y)=\frac{1}{2\sqrt{\pi}}(-2y)^{-1/4}e^{-\frac{2}{3}(-2y)^{3/2}}
\left(1+\mathcal{O}(|y|^{-3/4})\right)\underset{y\to-\infty}{\approx}0,
\ee whereas as $y\to +\infty$, it is described by
\be\label{asympnu0+}
\nu_0(y)\underset{y\to+\infty}{\approx}y^{1/2}\sum_{n=0}^{\infty}
\frac{b_n}{(2y)^{3n/2}}, \ee where $b_0=1$, $b_1=0$, and for $n
\geq 0$,
$$
b_{n+2}=4(9n^2-1)b_n-\frac{1}{2}\sum_{m=1}^{n+1}b_mb_{n+2-m}-\frac{1}{2}\sum_{l=1}^{n+1}\sum_{m=1}^{n+2-l}b_l
b_m b_{n+2-l-m}.
$$
\label{proposition-Painleve}
\end{prop}

Next, we construct $\nu_n\in H^\infty(\R)$ for $n\geq 1$ by induction on
$n$. For $n\geq 0$, we consider the following property:
\vspace{.2cm}\\
\null(H$_\text{n}$)\hfill $\forall k\in\{1,...,n\},$ \hfill $\left\{
\begin{minipage}{10.5cm}
$\bullet\ \nu_k\in H^\infty(\R)$ solves
(\ref{nun-eq}) (with $n$ replaced by $k$),\\
$\bullet\ \nu_k(y) \underset{y\to +\infty}{\approx}
y^{\beta-2k}\sum_{m=0}^{\infty} g_{k,m} y^{-3m/2}$ for some $\{ g_{k,m}\}_{m \in \mathbb{N}}$,\\
$\bullet$ $\nu_k(y) \underset{y\to -\infty}{\approx} 0$, \end{minipage}\right.$\hfill \vspace{.2cm}\\
where
$$\beta=\left\{\begin{array}{ll} -5/2&\text{if } d=1,\\ 1/2 &\text{if } d=2,3.\end{array}\right.$$
(H$_\text{0}$) is empty and, therefore, true by convention. Fix
$n\geq 1$ and assume that (H$_\text{n-1}$) is true. We are going
to construct $\nu_n$ such that (H$_\text{n}$) is satisfied. We
will make use of the following two lemmas, which are proved in
Sections 5 and 6.

\begin{lem}\label{asyop}
Let $W\in \mathcal{C}^1(\R)$ such that $W'\in L^\infty(\R_+)$ and there exists $C_0,C_+,A_+ > 0$ with
$$
W(x)\geq C_+ x \quad \mbox{\rm for} \quad x\geq A_+, \quad W(x)
\geq C_0 \quad \mbox{\rm for} \quad x\in \R,  \quad \mbox{\rm and}
\quad W'(x) \geq 0 \quad \mbox{\rm for} \quad x \geq A_+.
$$
Let $f\in L^2(\R)$ such that $x^\alpha f\in L^\infty(A_+,+\infty)$ for
some $\alpha>0$. Let
$$
\phi=(-\partial_x^2+W)^{-1}f\in H^1(\R).
$$
Then, as $x\to+\infty$, 
\be
\phi(x) &=& \mathcal{O}(x^{-(\alpha+1)}).\label{asympphi} 
\ee
Moreover, if $f$ and $W$ admit asymptotic series \be\label{serf}
f(x)\underset{x\to+\infty}{\approx}x^{-\alpha}\sum_{m=0}^{+\infty}c_mx^{-\gamma
  m},\quad W(x)\underset{x\to+\infty}{\approx}x\sum_{m=0}^{+\infty}v_mx^{-\gamma
  m},
\ee for some coefficients $\{ c_m \}_{m \in \mathbb{N}}$, $\{
v_m \}_{m \in \mathbb{N}}$ and $\gamma>0$ such that $3/\gamma$ is an integer, then $\phi$ admits an asymptotic
series \be\label{serphi}
\phi(x)\underset{x\to+\infty}{\approx}x^{-(\alpha+1)}\sum_{m=0}^{+\infty}d_mx^{-\gamma
m}, \ee for some coefficients $\{ d_m \}_{m \in \mathbb{N}}$. In
particular, as $x \to + \infty$, \be
\phi'(x)\ =\ \mathcal{O}(x^{-(\alpha+2)}),\quad
\phi''(x)\ =\ \mathcal{O}(x^{-(\alpha+3)}).\label{asympphi'2}
\ee
\end{lem}

\begin{lem}\label{V0below}
Let $W_0(y) := 3 \nu_0^2(y) - y$, where $\nu_0(y)$ is the solution of
the Painlev\'e-II equation (\ref{P2-eq}) given in Proposition
\ref{proposition-Painleve}. Then,
$$
W_{\rm min} := \underset{y\in\R}{\inf} W_0(y)>0.
$$
\end{lem}

>From the asymptotic behaviors of $\nu_0(y)$ as $y\to\pm\infty$, we
infer that
\be\label{W0inf}
W_0(y)\sim 2y \quad \mbox{\rm as} \quad  y\to +\infty \quad
\mbox{\rm and} \quad W_0(y)\sim -y \quad \mbox{\rm as} \quad y\to
-\infty.
\ee
Let us consider the operator
$$
M_0 := -4 \partial_y^2 + W_0(y)
$$
on $L^2(\R)$ with the domain,
$$
{\rm Dom}(M_0) = \{u\in L^2(\R) : \;\; -4 u'' + W_0 u \in
L^2(\R)\}.
$$
The Schr\"{o}dinger operator $M_0$ arises in the linearization of
the Painlev\'{e}-II equation at $\nu=\nu_0$. The
spectrum of $M_0$ is purely discrete and, thanks to Lemma
\ref{V0below}, it consists of a sequence of strictly positive
eigenvalues which goes to infinity. If $n=1$, it follows from the
choice of $\nu_0$ and from properties
(\ref{asympnu0-})-(\ref{asympnu0+}) that
\be\label{F1+}
F_1(y) \underset{y\to+\infty}{\approx}
(1-d)y^{-1/2}+y^{-7/2}\sum_{m=0}^{+\infty}3(m+2)b_{m+2}2^{-3(m+2)/2}(1-3(m+2))y^{-3m/2}
\ee
and
\be\label{F1-}
F_1(y) \underset{y\to -\infty}{\approx} 0.
\ee
Thanks to Lemma \ref{V0below}, we can look for $\nu_1$ solution of (\ref{nun-eq}) with $n=1$ in the form 
$$\nu_1(y)=\frac{(1-d)\Phi(y)}{W_0(y)y^{1/2}}+\tilde{\nu}_1(y),\quad y\in\R,$$
where $\Phi\in \mathcal{C}^\infty(\R)$ is such that $\Phi(y)\equiv 0$ if $y\leq 1/2$, $\Phi(y)\equiv 1$ if $y\geq 1$. Then, $\tilde{\nu}_1$ has to solve
\be\label{F1t}
-4\tilde{\nu}_1''(y)\!+\!W_0(y)\tilde{\nu}_1(y)=\tilde{F}_1(y):=F_1(y)-(1-d)y^{-1/2}\Phi(y)\!+\!4\frac{d^2}{dy^2}\left(\frac{(1-d)\Phi(y)}{W_0(y)y^{1/2}}\right)\!,\ y\in\R.\ \ 
\ee
>From the asymptotic expansions (\ref{asympnu0-})-(\ref{asympnu0+}) of $\nu_0$, we infer that $W_0$ also admits asymptotic expansions as $y\to\pm\infty$. Since  moreover $\nu_0\in\mathcal{C}^{\infty}(\R)$, it follows from (\ref{W0inf}), (\ref{F1+}), (\ref{F1-}) and (\ref{F1t}) that $\tilde{F}_1\in H^\infty(\R)$. Then, property (H$_\text{1}$) follows from Lemma \ref{asyop} applied on the one side to $\tilde{\nu}_1:=M_0^{-1}\tilde{F}_1$ with $\alpha = 7/2$, so that $\tilde{\nu}_1(y) = {\cal O}(y^{-9/2})$ as $y \to +\infty$, and on the other side to $y\mapsto\tilde{\nu}_1(-y)$ with $\alpha$ arbitrarily large. 
Furthermore, if $n\geq 2$, we have \be F_n(y)
&=&-\!\!\!\!\!\!\!\!\!\!\!\!\underset{\tiny{\begin{array}{c}0<n_1,n_2,n_3<n\\n_1+n_2+n_3=n\end{array}}}{\sum}\!\!\!\!\!\!\!\nu_{n_1}(y)\nu_{n_2}(y)\nu_{n_3}(y)
-3\!\!\!\!\!\!\!\!\underset{\tiny{\begin{array}{c}0<n_1,n_2<n\\n_1+n_2=n\end{array}}}{\sum}\!\!\!\!\nu_{0}(y)\nu_{n_1}(y)\nu_{n_2}(y)
-2d\nu_{n-1}'(y)-4y\nu_{n-1}''(y).\ \ \ \ \ \nonumber \ee Thanks to
(H$_\text{n-1}$), all the terms in the right hand side admit an
asymptotic expansion at $\pm\infty$. More precisely,
$$
F_n(y) \underset{y\to+\infty}{\approx}
y^{\beta+1-2n}\sum_{m=0}^{+\infty}f_{n,m}y^{-3m/2}
$$
for some coefficients $\{ f_{n,m} \}_{m \in \mathbb{N}}$, whereas
$$
F_n(y) \underset{y\to -\infty}{\approx}  0.
$$
Since $F_n\in\mathcal{C}^\infty(\R)$ and $n\geq 2$, we deduce that $F_n\in H^\infty(\R)$,
and we can define $\nu_n = M_0^{-1}F_n\in
H^\infty(\R)$. By Lemma \ref{asyop} with $\gamma=3/2$ and $\alpha=2n-\beta-1$, we then have
$$
\nu_n(y) \underset{y\to+\infty}{\approx}
y^{\beta-2n}\sum_{m=0}^{+\infty}g_{n,m}y^{-3m/2}
$$
for some coefficients $\{ g_{n,m} \}_{m \in \mathbb{N}}$, and
$$
\nu_n(y) \underset{y\to-\infty}{\approx}  0
$$
where we have applied Lemma \ref{asyop} with $\gamma=3/2$ to the
function $\nu_n(-y)$. Therefore, (H$_\text{n}$) is true, which completes the
construction by induction of the sequence of
solutions $\{ \nu_n(y) \}_{n \geq 1}$ of the inhomogeneous equations (\ref{nun-eq}).

\subsection{Construction of $R_{N,\eps}$}
In this subsection, we construct a solution
$R_{N,\eps}$ to equation (\ref{r-eq}), such that given the $\nu_n$'s
constructed in subsection 2.1, expansion (\ref{exp}) provides
a solution of equation (\ref{nu-eq}). The solution $R_{N,\eps}$ of
equation (\ref{r-eq}) is obtained by a fixed point argument. In order
to explain the functional framework in which the fixed point theorem
will be applied, let us first introduce the functional spaces
$$
L^2_\eps=\left\{u\in L^1_{\text{loc}}(J_{\eps}) : \quad (1-\eps^{2/3} y)^{d/4-1/2}u \in L^2(J_{\eps})\right\}
$$
and
$$
H^1_\eps=\left\{u\in L^2_\eps : \quad (1-\eps^{2/3}y)^{d/4} u' \in
L^2(J_{\eps})\text{ and }(1-\eps^{2/3} y)^{d/4-1/2}W_0^{1/2}u \in L^2(J_{\eps}) \right\},
$$
endowed with their respective squared norms
$$
\| u \|^2_\eps:=\| u \|^2_{L^2_\eps} = \int_{-\infty}^{\eps^{-2/3}} (1-\eps^{2/3} y)^{d/2-1}u^2 dy
$$
and
$$
\| u\|^2_{H^1_\eps} := \int_{-\infty}^{\eps^{-2/3}} \left[4 (1 - \eps^{2/3}
y)^{d/2} |u'|^2+(1-\eps^{2/3} y)^{d/2-1}W_0 u^2  \right] dy.
$$
We are looking for a solution $R_{N,\eps}(y)$ of
Equation (\ref{r-eq}) on $J_\eps$ such that the function $R_{N,\eps}(\eps^{-2/3}(1-|x|^2))$ is regular on $\R^d$. As a result, it is convenient for the
sequel to introduce the map $T^\eps:L_\eps^2\mapsto L_r^2(\R^d)$ defined for $u\in L_\eps^2$ by
$$(T^\eps u)(z):=u(\eps^{-2/3}-\eps^{2/3}|z|^2),$$
which makes the link between functions defined on $J_\eps$ and radial
functions defined on $\R^d$, in terms of the variable $z=\eps^{-2/3}x\in\R^d$. An easy calculation shows that $T^\eps$ is a bijection from $L_\eps^2$ into $L_r^2(\R^d)$, and that for every $u\in L_\eps^2$, 
\be\label{teul}
\|T^\eps u\|_{L^2(\R^d)}^2 & = & \frac{|\mathbb{S}^{d-1}|}{2\eps^{2(d-1)/3}}\|u\|_\eps^2.
\ee
Moreover, $T^\eps$ induces a bijection from $H_\eps^1$ into
$$Q_\eps:=\left\{u\in L^2(\R^d): \int_{\R^d}\left[|\nabla u|^2+W_0(\eps^{-2/3}-\eps^{2/3}|z|^2)|u|^2\right]dz<\infty\right\}$$
and for every $u\in H^1_\eps$,
\be\label{teuh}
\|T^\eps u\|_{Q_\eps}^2 & = &\int_{\R^d}\left[|\nabla T^\eps u|^2+W_0(\eps^{-2/3}-\eps^{2/3}|z|^2)|T^\eps u|^2\right]dz=\frac{|\mathbb{S}^{d-1}|}{2\eps^{2(d-1)/3}}\|u\|_{H^1_\eps}^2.
\ee

Let us rewrite equation (\ref{r-eq}) for the remainder term
$R_{N,\eps}(y)$ in the operator form
\begin{equation}
\label{operator-form} M^\eps R_{N,\eps}(y) =
F_{N,\eps}(y,R_{N,\eps}),\quad y
\in J_{\eps},
\end{equation}
where $M^\eps$ is the self-adjoint operator on $L^2_\eps$ defined by
\begin{equation}
\label{leps} 
\left\{
\begin{array}{l}
{M}^\eps := -4 (1-\eps^{2/3}y)^{-d/2+1} \partial_y
(1-\eps^{2/3}y)^{d/2} \partial_y +
W_{0}(y)=(T^\eps)^{-1}K^\eps T^\eps, \\
{\rm Dom}({M}^\eps) = \left\{u\in
  L^2_\eps : \;\;
  K^\eps T_{\eps} u\in L^2(\R^d) \right\}
  \end{array} \right.
\end{equation}
and $K^\eps$ denotes the Schr\"odinger operator on $L^2(\R^d)$, 
$$K^\eps:=-\Delta+W_0(\eps^{-2/3}-\eps^{2/3}|z|^2).$$
The solution $R_{N,\eps}$ of the nonlinear equation
(\ref{operator-form}) will be obtained from the fixed point theorem
applied to the map
$$
\Phi_{N,\eps}:R\mapsto (M^\eps)^{-1}F_{N,\eps}(\cdot,R),
$$
which will be shown to be continuous from $H^1_\eps$ into
itself. First, we shall prove the following lemma.

\begin{lem}
\label{lemma-map} The operator $M^\eps$ is invertible, and for every
$f\in L^2_\eps$, 
$$
\| (M^\eps)^{-1}f \|_{H^1_{\eps}} \leq W_{\min}^{-1/2}\| f \|_\eps.
$$
\end{lem}

\begin{Proof}
Let us consider the continuous,
bilinear, coercive form on $Q_\eps$ defined by
$$
a(u,v) = \int_{\R^d} \left[ \nabla u\nabla v+W_0(\eps^{-2/3}-\eps^{2/3} |z|^2)uv \right] dz.
$$
By the Cauchy-Schwarz inequality and Lemma \ref{V0below}, for every $f\in L_\eps^2$,
$$
v \mapsto \int_{\R^d}T^\eps fvdz
$$
defines a continuous linear form on $Q_\eps$. Thus, by the Lax-Milgram
Theorem \cite{T}, there exists a unique $\psi\in Q_\eps$ such that for
every $v\in Q_\eps$,
$$
a(\psi,v)=\int_{\R^d}T^\eps fvdz.
$$
Moreover, $\psi \in Q_{\eps}$ is radial and satisfies
$$
 K^\eps \psi = T^\eps f \quad
\mbox{\rm in} \quad \mathcal{D}'(\R^d).
$$
Thus, $\phi:=(T^\eps)^{-1}\psi\in H_\eps^1\cap {\rm Dom}(M^\eps)$ satisfies
$$M^\eps \phi=f.$$
>From (\ref{teuh}) and a calculation similar to (\ref{teul}), we also check that
\begin{eqnarray*}
\|\phi\|_{H_\eps^1}^2 &= &\frac{2\eps^{2(d-1)/3}}{|\mathbb{S}^{d-1}|}a(\psi,\psi) =\frac{2\eps^{2(d-1)/3}}{|\mathbb{S}^{d-1}|}\int_{\R^d}T^\eps f\psi dz=\int_{-\infty}^{\eps^{-2/3}}(1-\eps^{2/3}y)^{d/2-1}f\phi dy\\
&\leq &\|f\|_\eps
\|\phi\|_\eps\leq W_{\text{min}}^{-1/2}\|f\|_\eps\|\phi\|_{H_\eps^1},
\end{eqnarray*}
from which the upper bound on $\phi=(M^\eps)^{-1}f$ in $H^1_{\eps}$ follows.
\end{Proof}

Next, we prove that
$R\mapsto F_{N,\eps}(\cdot,R)$ continuously maps $H_\eps^1$ into
$L_\eps^2$. We write
$$
F_{N,\eps}(y,R)=F_{N,0}(y)+G_{N,\eps}(y,R),
$$
where
\be
F_{N,0}&=&-(4y\nu_N''+2d\nu_N')-\underset{\tiny{\begin{array}{c}n_1+n_2+n_3=N+1\\0\leq
      n_1,n_2,n_3\leq
      N\end{array}}}{\sum}\nu_{n_1}\nu_{n_2}\nu_{n_3}
\ee
and
\be\label{gn-eq}
G_{N,\eps}&=&-\sum_{n=1}^{2N-1}\eps^{2n/3}\underset{\tiny{\begin{array}{c}n_1+n_2+n_3=n+N+1\\0\leq
      n_1,n_2,n_3\leq
      N\end{array}}}{\sum}\nu_{n_1}\nu_{n_2}\nu_{n_3}-\left(3\sum_{n=1}^{2N}\eps^{2n/3}\underset{\tiny{\begin{array}{c}n_1+n_2=n\\0\leq
      n_1,n_2\leq N\end{array}}}{\sum}\nu_{n_1}\nu_{n_2}\right)R\nonumber\\
&&-\left(3\sum_{n=N+1}^{2N+1}\eps^{2n/3}\nu_{n-(N+1)}\right)R^2-\eps^{4(N+1)/3}R^3.
\ee
We first show that $F_{N,0}\in L_\eps^2$. Indeed, from the properties
of the $\nu_n$'s, we infer that
$$F_{N,0}(y) \underset{y\to -\infty}{\approx} 0,$$
and $F_{N,0}$ also admits an asymptotic expansion as $y\to +\infty$, with
$$
4y\nu_N''(y)+2d\nu_N'(y) = {\cal O}(y^{\beta-1-2N})
$$
and if $n_1+n_2+n_3=N+1$,
$$
\nu_{n_1}\nu_{n_2}\nu_{n_3}(y) = \left\{\begin{array}{ll} {\cal O}(y^{-19/2-2N})&
\text{ if } d=1\text{ and }n_1,n_2,n_3>0,\\ {\cal O}(y^{-13/2-2N})&
\text{ if } d=1\text{ and }n_1 \text{ or }n_2\text{ or }n_3=0, \\
     {\cal O}(y^{-1/2-2N})&\text{ if } d\geq 2,\end{array}\right.
$$
(notice that $n_1+n_2+n_3=N+1$ with $0\leq
      n_1,n_2,n_3\leq N$ implies that at most one of the numbers
      $n_1,n_2,n_3$ is equal to 0). Since $N\geq 2$, we deduce that in any case,
$$
F_{N,0}(y) = {\cal O}(y^{-9/2})\   \mbox{\rm as $y\to+\infty$,\quad while} \quad
F_{N,0}(y) \underset{y\to -\infty}{\approx} 0.
$$
Therefore, for $\alpha>0$ sufficiently large and $\eps<1$,
\begin{eqnarray*}
\nonumber \int_{-\infty}^{\eps^{-2/3}}
(1-\eps^{2/3}y)^{d/2-1}F^2_{N,0}dy & \lesssim &
\int_{-\infty}^1(1+|y|)^{-2\alpha}dy+\int_1^{\eps^{-2/3}}y^{-9}(1-\eps^{2/3}y)^{d/2-1}dy.
\end{eqnarray*}
In the case $d=1$, the second integral in the right hand side is
estimated by
\begin{eqnarray*}
\int_1^{\eps^{-2/3}}y^{-9}(1-\eps^{2/3}y)^{-1/2}dy& \lesssim &
\eps^{16/3}\int_{\eps^{2/3}}^1\frac{z^{-9}}{(1-z)^{1/2}}dz \\
\nonumber &\lesssim &
\eps^{16/3}\sqrt{2}\int_{\eps^{2/3}}^{1/2}z^{-9}dz+\frac{\eps^{16/3}}{2^9}\int_{1/2}^1
\frac{1}{(1-z)^{1/2}}dz \lesssim   1,
\end{eqnarray*}
whereas for $d\geq 2$,
\begin{eqnarray*}
\int_1^{\eps^{-2/3}}y^{-9}(1-\eps^{2/3}y)^{d/2-1}dy& \lesssim &
\int_1^{\eps^{-2/3}}y^{-9}dy\lesssim 1.
\end{eqnarray*}
Therefore in both cases $F_{N,0}\in L^2_{\eps}$ and
\begin{eqnarray}\label{cst}
\| F_{N,0} \|_{\eps}& \lesssim &  1.
\end{eqnarray}
Similarly, the term which does not depend on $R$ in the right hand
side of (\ref{gn-eq}) is ${\cal O}_{L_\eps^2}(\eps^{2/3})$. Let
$R\in H_\eps^1$. To estimate the linear term in $R$ in the
definition of $G_{N,\eps}$, notice that if $n_1+n_2=n\geq 1$, then
$n_1$ or $n_2$ is not equal to 0, thus $\nu_{n_1}\nu_{n_2}(y) =
{\cal O}(y^{-1})$ as $y\to +\infty$. In particular,
$\nu_{n_1}\nu_{n_2}\in L^\infty(\R)$ and \be\label{lin}
\|\nu_{n_1}\nu_{n_2} R\|_\eps\leq \|\nu_{n_1}\nu_{n_2}
\|_{L^\infty(\R)}\|R\|_\eps\leq\|\nu_{n_1}\nu_{n_2}
\|_{L^\infty(\R)}W_{\text{min}}^{-1/2}\|R\|_{H^1_\eps}\lesssim
\|R\|_{H^1_\eps}. \ee
In order to estimate the quadratic and cubic terms in the right hand
side of (\ref{gn-eq}), the following lemma will be useful.

\begin{lem}\label{emb}
Let $p=1,2$ or $3$. There
exists a $\eps$-independent constant $C>0$ such
that for every $\eps>0$, if $u\in H_\eps^1$ then $u^p\in L_\eps^2$
and 
$$
\|u^p\|_{\eps} \leq C\eps^{-(p-1)(d-1)/3}\|u\|_{H_\eps^1}^p.
$$
\end{lem}

\begin{Proof}
Let $u\in H_\eps^1$. We have checked in (\ref{teuh}) that $T^\eps u\in Q_\eps\subset H^{1}(\R^d)$ and
\be\label{B}
\|T^\eps u\|_{H^{1}(\R^d)} \lesssim \|T^\eps u\|_{Q_\eps} 
&\lesssim & \eps^{-(d-1)/3}\|u\|_{H_\eps^1}.
\ee
By Sobolev embeddings, it follows that $T^\eps u\in L^{2p}(\R^d)$, and
\begin{eqnarray}\label{A}
\|u^p\|_{\eps}&\lesssim &\eps^{(d-1)/3}\|T^\eps(u^p)\|_{L^2(\R^d)}= \eps^{(d-1)/3}\|T^\eps u\|_{L^{2p}(\R^d)}^{p}\nonumber\\
&\lesssim &\eps^{(d-1)/3}\|T^\eps u\|_{H^{1}(\R^d)}^{p}\lesssim \eps^{-(p-1)(d-1)/3}\|u\|_{H_\eps^1}^p,
\end{eqnarray}
where we have also made use of (\ref{teul}) with $u$ replaced by $u^p$. 
\end{Proof}

\begin{rem} The statement of Lemma \ref{emb} can be extended
for all values of $p$ for which $H^1(\R^d)$ is continuously embedded
into $L^{2p}(\R^d)$, that is $1\leq p\leq \infty$ for $d=1$, $1\leq p<
\infty$ for $d=2$ and  $1\leq p\leq 3$ for $d=3$. 
\end{rem}

Thanks to Lemma \ref{emb}, for any integer $k\geq 1$,
\be\label{quadk1} \|\nu_k
R^2\|_\eps\lesssim\|\nu_k\|_{L^\infty(\R)}\eps^{-(d-1)/3}\|R\|_{H_\eps^1}^2\lesssim\eps^{-(d-1)/3}\|R\|_{H^1_{\eps}}^2,
\ee whereas for $k=0$, 
\be\label{quadk0} \|\nu_0
R^2\|_\eps\lesssim
\|\nu_0\|_{L^\infty(J_{\eps})}  \eps^{-(d-1)/3}   \|R\|_{H_\eps^1}^2 \lesssim
\eps^{-d/3}\|R\|_{H_\eps^1}^2 \ee 
On the other side, 
\be\label{cub}
\|R^3\|_\eps \lesssim\eps^{-2(d-1)/3}\|R\|_{H_\eps^1}^3, 
\ee 
thanks to
Lemma \ref{emb} again. By Lemma \ref{lemma-map} as well as bounds
(\ref{cst}), (\ref{lin}), (\ref{quadk1}), (\ref{quadk0}) and
(\ref{cub}), 
$$\|\Phi_{N,\eps}(R)-R_{N,\eps}^0\|_{H_\eps^1}\lesssim\eps^{2/3}+\eps^{2/3}\|R\|_{H_\eps^1} + \eps^{(2N+2-d)/3}
\|R\|_{H_\eps^1}^2+\eps^{(4N+6-2d)/3}\|R\|_{H_\eps^1}^3,$$ where
$$
R_{N,\eps}^0 :=
(M^\eps)^{-1}F_{N,0}.
$$
In particular, for $\eps > 0$ sufficiently small and for some $\eps$-independent constant $C>0$,  $\Phi_{N,\eps}$ maps the
ball
$$
B_\eps:=B_{H_\eps^1}(R_{N,\eps}^0,C\eps^{2/3})
$$
into itself, where we have used the assumption $N\geq 2$. Similarly, there exists an $\eps$-independent constant
$\tilde{C}>0$ such that for every $R_1,R_2$
in $B_\eps$,
$$\|\Phi_{N,\eps}(R_1)-\Phi_{N,\eps}(R_2)\|_{H_\eps^1}\leq \tilde{C}\eps^{2/3}\|R_1-R_2\|_{H_\eps^1}.$$
As a result, provided $\eps$ is sufficiently small, $\Phi_{N,\eps}$ is a
contraction on $B_\eps$. The Fixed Point
Theorem ensures that $\Phi_{N,\eps}$ has a unique fixed point $R_{N,\eps}\in
B_\eps$. In particular,
\be\label{ball}
\|R_{N,\eps}-R_{N,\eps}^0\|_{H_\eps^1}\lesssim
\eps^{2/3}.
\ee
We next prove that $R_{N,\eps}$ satisfies the regularity properties stated in Theorem \ref{theorem-main}. The fixed point $R_{N,\eps}\in H^1_\eps$ of $\Phi_{N,\eps}$ has been constructed in such a way that $T^\eps R_{N,\eps}\in H^1(\R^d)$ solves the equation
\be\label{ltr}
K^\eps T^\eps R_{N,\eps}=T^\eps (F_{N,\eps}(\cdot, R_{N,\eps}))\in L^2(\R^d).
\ee
Thanks
to Lemma \ref{lemma-map} and (\ref{cst}),  we obtain
\be\label{rnli}
\| R_{N,\eps}^0 \|_{H^1_{\eps}} \lesssim
\| F_{N,0} \|_\eps \lesssim 1.
\ee
Thus, (\ref{ball}) yields
\be\label{rfordom}
\| R_{N,\eps} \|_\eps\lesssim\| R_{N,\eps} \|_{H^1_{\eps}} \lesssim 1.
\ee
As a result, from (\ref{cst}), (\ref{lin}), (\ref{quadk1}), (\ref{quadk0}) and
(\ref{cub}), we infer
\be\label{mer}
\|F_{N,\eps}(\cdot,R_{N,\eps})\|_\eps\lesssim 1.
\ee
>From (\ref{ltr}), (\ref{mer}) and (\ref{teul}) we deduce
\be
\|K^\eps T^\eps R_{N,\eps}\|_{L^2(\R^d)}=\|T^\eps (F_{N,\eps}(\cdot, R_{N,\eps}))\|_{L^2(\R^d)}\lesssim \eps^{-(d-1)/3}.
\ee
Next, we use the following Lemma, which is proved in Section 7.
\begin{lem}\label{shen}
$(K^\eps)^{-1}\in \mathcal{L}(L^2(\R^d), H^2(\R^d))$ is
uniformly bounded in $\eps$.
\end{lem}
As a result, we infer from the Sobolev embedding of $H^2(\R^d)$ into
$L^\infty(\R^d)$ that $T^\eps R_{N,\eps}\in H^2(\R^d)$ and 
\be\label{rnelinf}
\|R_{N,\eps}\|_{L^\infty(J_\eps)}=\|T^\eps R_{N,\eps}\|_{L^\infty(\R^d)}\lesssim\eps^{-(d-1)/3}.
\ee
Moreover, a bootstrapping argument shows that $T^\eps
R_{N,\eps}\in\mathcal{C}^\infty(\R^d)$. As a result, $R_{N,\eps}\in\mathcal{C}^\infty(J_\eps)$.

\subsection{$\nu_\eps(y)>0$ for all $y \in J_{\eps}$}

We have constructed above $\{\nu_n\}_{n\geq 0}$ and $R_{N,\eps}$ in such a way that 
\be\label{tetae}
\tilde{\eta}_\eps(x)&:=&\eps^{1/3}\sum_{n=0}^N\eps^{2n/3}\nu_n\left(\frac{1-|x|^2}{\eps^{2/3}}\right)+\eps^{2N/3+1}(T^\eps
R_{N,\eps})(\eps^{-2/3}x)\nonumber\\
&=&\eps^{1/3}\nu_\eps\left(\frac{1-|x|^2}{\eps^{2/3}}\right),\quad
x\in\R^d 
\ee
is a classical, radially symmetric solution of equation (\ref{eta-eq}).
In order to
claim that $\tilde{\eta}_\eps$ is a ground state, it is sufficient to check that
$\tilde{\eta}_\eps(x)>0$, for every $x\in\R^d$, which is equivalent to
$\nu_\eps(y)>0$, for every $y \in J_{\eps}$. 

For every $n \geq 1$, $\| \nu_n \|_{L^{\infty}(\R)} \lesssim 1$. Therefore, from (\ref{rnelinf}), (\ref{tetae}), since $N\geq 2$, we deduce the existence of a
constant $C>0$ such that for every $y\in J_{\eps}$,
$$
\nu_\eps(y)-\nu_0(y)\geq -C\eps^{2/3}.
$$
Since $\nu_0(y)$ increases from $0$ to $+\infty$ as $y$ goes from
$-\infty$ to $+\infty$, we deduce that for $\eps\ll 1$,
$$
\nu_\eps(y)\geq\nu_0(-1) -C\eps^{2/3}>0,  \quad y \in [-1,\eps^{-2/3}].
$$
Coming back to the variable $x$, it follows that
\be\label{tet>0}
\tilde{\eta}_\eps(x)>0,\quad |x|\leq (1+\eps^{2/3})^{1/2}.
\ee
It remains to prove that $\tilde{\eta}_{\eps}(x) > 0$ for all $|x| > (1+\eps^{2/3})^{1/2}$.
Assume by contradiction that $\tilde{\eta}_\eps$ is not strictly
positive on $\R^d$. Then, let
$$
r_\eps=\inf\{r>0,\ \tilde{\eta}_\eps(r)=0\} \in \left( (1+\eps^{2/3})^{1/2},\infty\right)
$$
where for convenience, since $\tilde{\eta}_\eps$ is radial, we denote $\tilde{\eta}_\eps(|x|)=\tilde{\eta}_\eps(x)$. By construction, $\tilde{\eta}_\eps(r_\eps)=0$ and
$\tilde{\eta}_\eps'(r_\eps)\leq 0$. If $\tilde{\eta}_\eps'(r_\eps)=
0$, then $\tilde{\eta}_\eps \equiv 0$, because $\tilde{\eta}_\eps(r)$
satisfies the differential equation
$$
-\frac{1}{r^{d-1}}\frac{d}{dr}\left(r^{d-1}\frac{d}{dr}\tilde{\eta}_\eps\right)(r)+ \frac{1}{\eps^2}(r^2-1+\tilde{\eta}_\eps(r)^2) \tilde{\eta}_\eps(r)=0.
$$ 
This is a contradiction with
(\ref{tet>0}). Thus, $\tilde{\eta}_\eps'(r_\eps)< 0$. Let
$$
\tilde{r}_\eps := \sup\{r>r_\eps,\tilde{\eta}_\eps(r')<0\text{ for
} r' \in (r_\eps, r)\}\in (r_\eps,+\infty].
$$
Then, for every $r\in (r_\eps, \tilde{r}_\eps)$,
$$
\frac{d}{dr}\left(r^{d-1}\frac{d}{dr}\tilde{\eta}_\eps\right)(r)= \frac{r^{d-1}}{\eps^2}(r^2-1+\tilde{\eta}_\eps(r)^2) \tilde{\eta}_\eps(r)\leq
0,
$$
and we deduce by integration that for every $r\in (r_\eps, \tilde{r}_\eps)$,
$$
r^{d-1}\tilde{\eta}_\eps'(r) \leq r_\eps^{d-1}\tilde{\eta}_\eps'(r_\eps)<0,
$$
and
\be\label{limalinf}
\tilde{\eta}_\eps(r)\leq
r_\eps^{d-1}\tilde{\eta}_\eps'(r_\eps)\int_{r_\eps}^r s^{1-d}ds.
\ee
The right hand side in (\ref{limalinf}) is a negative, decreasing
function of $r$, which implies $\tilde{r}_\eps = +\infty$, as well as
a contradiction with the fact that $\tilde{\eta}_\eps(r)\to 0$ as $r\to
+\infty$. Therefore $\tilde{\eta}_{\eps}(r)>0$ for all $r \in \mathbb{R}_+$.

\section{Spectrum of the Schr\"odinger operator $L_+^\eps$ in
  the case $d=1$}

Consider the Schr\"odinger operator
$$
L_+^{\eps} = -\eps^2 \partial_x^2 + V_{\eps}(x), \quad
V_{\eps}(x) = 3 \eta^2_{\eps}(x) - 1 + x^2,
$$
associated with the stationary Gross--Pitaevskii equation (\ref{stationaryGP})
linearized at the ground state $\eta_{\eps}$. It is a self-adjoint
operator on $L^2(\R)$. Since the potential $V_{\eps}(x)$
is confining in the sense of $V_{\eps}(x) \to +\infty$ as $|x| \to
\infty$, $L_+^{\eps}$ has compact resolvent and a purely discrete
spectrum. By Sturm-Liouville theory, the
eigenvalues of $L_+^{\eps}$, denoted $\{ \lambda_n^\eps \}_{n \geq 1}$
(sorted in increasing order) are simple. Moreover,
thanks to the even symmetry of $V_{\eps}$ on $\R$, the eigenfunctions
of $L_+^{\eps}$ corresponding to $\lambda_n^\eps$ are even
(resp. odd) in $x$ if $n$ is odd (resp. even). If $\lambda$ is an
eigenvalue of $L_+^\eps$ and $\phi\in L^2(\R)$ is a corresponding
eigenfunction, we define a function $v\in
L_\eps^2$ by
$$
\phi(x)=v\left(\frac{1-x^2}{\eps^{2/3}}\right), \quad  x \in \R_+.
$$
Let us denote
$W_{\eps}(y) = 3 \nu^2_{\eps}(y) - y$. Then, $\phi\in L^2(\R)$ is an
even eigenfunction of $L_+^{\eps}$ corresponding to the eigenvalue
$\lambda$ if and only if $v\in L^2_\eps$ satisfies
the differential equation
\begin{equation}
\label{ut}  \left(-4 (1-\eps^{2/3}y)^{1/2}\partial_y
(1-\eps^{2/3}y)^{1/2}
\partial_y + W_{\eps}(y)\right)v(y)=\eps^{-2/3}\lambda v(y),\quad y\in J_\eps
\end{equation}
and the Neumann boundary condition\\
\null\hfill
$\phi'(0)=- 2\eps^{-2/3}\left((1-\eps^{2/3}y)^{1/2} v'(y)\right) \biggr|_{y=\eps^{-2/3}} =0$.\hfill (NC)\vspace{.1cm}\\
Similarly, $\phi\in L^2(\R)$ is an odd eigenfunction of
$L_+^{\eps}$ corresponding to the eigenvalue $\lambda$ if and only
if $v\in L^2_\eps$ satisfies (\ref{ut}) and the
Dirichlet
boundary condition\vspace{.1cm}\\
\null\hfill
$\phi(0)=v(\eps^{-2/3})=0$.\hfill (DC)\vspace{.1cm}\\
As a result, the eigenvalues of $L_+^{\eps} $ are directly related to
the eigenvalues of the two self-adjoint operators on $L^2_\eps$,
$\check{M}^\eps$ and $\tilde{M}^\eps$, where
$$
\left\{
\begin{array}{l}
\check{M}^\eps := -4 (1-\eps^{2/3}y)^{1/2} \partial_y
(1-\eps^{2/3}y)^{1/2} \partial_y +
W_{\eps}(y), \\
{\rm Dom}(\check{M}^\eps) = \left\{v\in
  L^2_\eps : \;\;
  \check{M}^{\eps} v\in L^2_\eps \text{ and } v\text{ satisfies (NC)} \right\},
  \end{array} \right.
$$
and $\tilde{M}^\eps$ is defined similarly by replacing (NC) by (DC)
in the definition of the domain. Namely, if we denote
$\{\check{\mu}_n^\eps\}_{n\geq 1}$ (resp.
$\{\tilde{\mu}_n^\eps\}_{n\geq 1}$) the eigenvalues of
$\check{M}^\eps$ (resp. $\tilde{M}^\eps$) sorted in increasing
order, then for every $n\geq 1$,
$$\check{\mu}_n^\eps=\eps^{-2/3}\lambda_{2n-1}^\eps\quad
\text{and}\quad \tilde{\mu}_n^\eps=\eps^{-2/3}\lambda_{2n}^\eps.$$
As $\eps\to 0$, the eigenvalue problems (\ref{ut})
for the operators $\check{M}^\eps$ and $\tilde{M}^\eps$ formally
converge to the eigenvalue problem for the
Schr\"{o}dinger operator $M_0$ defined after Lemma \ref{V0below},
$$
(-4\partial_y^2 +W_0(y))v(y)=\mu v(y), \quad y\in \R,\quad \text{where }\mu=\eps^{-2/3}\lambda .
$$
By the discussion below
Lemma \ref{V0below}, the purely discrete spectrum of $M_0$ in $L^2(\R)$ consists of
an increasing sequence of positive eigenvalues $\{ \mu_n \}_{n \geq
  1}$. We shall prove that the
eigenvalues  of $L_+^{\eps}$ converge to the
eigenvalues of $M_0$ as $\eps\to 0$, according to the following result.

\begin{theo}
\label{theorem-eigenvalues}
The spectrum of $L_+^{\eps}$ consists
of an increasing sequence of positive eigenvalues $\{ \lambda_n^\eps
\}_{n \geq 1}$ such that for each $n \geq 1$,
\begin{equation}
\label{asymptotic-limit-eigenvalues}
\lim_{\eps \downarrow 0} \frac{\lambda_{2n-1}^\eps}{\eps^{2/3}} = \lim_{\eps \downarrow 0} \frac{\lambda_{2n}^\eps}{\eps^{2/3}} = \mu_n.
\end{equation}
\end{theo}

\begin{Proof}
We prove only the convergence of 
$\tilde{\mu}_n^\eps =\lambda_{2n}^\eps/\eps^{2/3}$ to $\mu_n$, for every $n\geq 1$. The proof of the
convergence of $\check{\mu}_n^\eps
=\lambda_{2n-1}^\eps/\eps^{2/3}$ to $\mu_n$ is identical.

Denote by $\left<\cdot,\cdot\right>$ and $\|\cdot\|$ the scalar
product and the norm in $L^2(\R)$, and by
$\left<\cdot,\cdot\right>_\eps$ and $\|\cdot\|_\eps$ the scalar
product and the norm in $L^2_\eps$. If $u,v\in
L^2(\R)$, $u\perp v$ means that $\left< u,v\right>=0$, whereas if
$u,v\in L^2_\eps$, $u\perp_\eps v$ means that
$\left< u,v\right>_\eps=0$. We denote by
$$\rqe(v)=\frac{\fqe(v,v)}{\|v\|_{\eps}^2}$$
the Rayleigh quotient for the operator $\met$, where $\qet$ denotes
the corresponding bilinear form
$$\qet(u,v)=\int_{J_\eps}\left(4 (1-\eps^{2/3}y)^{1/2} \partial_y
  u\partial_y
  v+\frac{W_\eps(y)}{(1-\eps^{2/3}y)^{1/2}}u(y)v(y)\right)dy,$$
defined for $u,v\in H^1_{\eps}$. Similarly,
$$R(v)=\frac{Q(v,v)}{\|v\|^2}$$ denotes the Rayleigh quotient for $M_0$, where $Q$ is
the corresponding bilinear form
$$Q(u,v)=\int_{\R}\left(4\partial_y
  u\partial_y
  v+W_0(y)u(y)v(y)\right)dy,$$
defined for  $u,v\in\{u\in H^1(\R):W_0^{1/2} u\in
L^2(\R)\}$. 

Let $\tilde{u}_n^\eps$ (resp. $u_n$) denote an eigenfunction
of $\met$ (resp. $M_0$) corresponding to the eigenvalue
$\tilde{\mu}_n^\eps$ (resp. $\mu_n$), normalized by $\|\tilde{u}_n^\eps\|_\eps=1$ (resp. $\|u_n\|=1$).The
eigenvalues of $M_0$ are given by the Max-Min principle:
\be\label{ray}
\mu_n=\underset{\begin{array}{c}v\in {\rm Dom}(M_0)\\ v\perp
    u_1,\cdots,u_{n-1}\end{array}}{\rm inf}R(v),
\ee
whereas the eigenvalues of $\met$ are similarly given by
\be\label{rayt}
\mne=\underset{\begin{array}{c}v\in {\rm Dom}(\met)\\ v\perp_\eps
    \tilde{u}_1^\eps,\cdots,\tilde{u}_{n-1}^\eps\end{array}}{\rm
  inf}R^\eps(v).
\ee
Let us fix $\delta\in (0,2/3)$. Let $\Phi\in \mathcal{C}^\infty(\R)$
be an non-decreasing function such that $\Phi\equiv 0$ on $\R_-$ and
$\Phi\equiv 1$ on $[1,+\infty)$. For $\eps>0$ sufficiently small, we
also define $\chi_\eps\in\mathcal{C}_c^\infty(\R)$ by
$$\chi_\eps(x)=\Phi\left(\frac{2x+\eps^{-2/3}}{\eps^{-2/3}-2\eps^{-\delta}}\right)\Phi\left(\frac{\eps^{-2/3}-2x}{\eps^{-2/3}-2\eps^{-\delta}}\right),$$
such that $\chi_\eps$ is even, $\chi_\eps\equiv 1$ on
$[-\eps^{-\delta},\eps^{-\delta}]$ and $Supp(\chi_\eps)\subset
[-\frac{\eps^{-2/3}}{2},\frac{\eps^{-2/3}}{2}]$.  We shall prove
recursively the
following properties:
\vspace{.2cm}\\
\null(G$_\text{n}$)\hfill $\left\{
\begin{minipage}{10.5cm}\begin{enumerate}
\item $\tilde{\mu}_n^\eps=\mu_n+\mathcal{O}(\eps^{2/3-\delta})$,
\item for every $k\geq n+1$,
$\left<\chi_\eps u_k,\tilde{u}_n^\eps\right>_\eps=\mathcal{O}(\eps^{1/3-\delta/2}),$
\item for every $k\geq n$,
$\left<\chi_\eps\tilde{u}_k^\eps,{u}_{n-1}\right>=\mathcal{O}(\eps^{1/3-\delta/2}),$
\item $\underset{c\in\R}{\inf}\|\chi_\eps\tilde{u}_n^\eps-c u_n\|=\mathcal{O}(\eps^{1/3-\delta/2})$,
\item $\underset{c\in\R}{\inf}\|\chi_\eps u_{n-1}-c\tilde{u}_{n-1}^\eps\|_\eps=\mathcal{O}(\eps^{1/3-\delta/2})$,\end{enumerate}\end{minipage}\right.$\hfill
\vspace{.2cm}\\
where for $n=1$, (iii)$_1$ and (v)$_1$ have to be understood as empty properties. Let us fix $n\geq 1$ and
assume that (G$_\text{k}$) is true for every
$k\in \{1,\cdots ,n-1\}$ (for $n=1$, this condition is
empty, therefore true by convention). The proof of
(G$_\text{n}$) is then divided in five steps.

\vspace{0.2cm}

{\bf Step 1. Upper bound on $\tilde{\mu}_n^\eps$.} First, we shall
prove that
\be\label{v}
R^\eps\left(v_n^\eps\right)=\mu_n+\mathcal{O}(\eps^{2/3-\delta}),
&where&
v_n^\eps=\chi_\eps u_n-\sum_{k=1}^{n-1}\left<\chi_\eps u_n,
    \tilde{u}_k^\eps\right>_\eps\tilde{u}_k^\eps.
\ee Then, thanks to (\ref{rayt}), since  $v_n^\eps\in {\rm
Span}(\tilde{u}_1^\eps,\cdots
,\tilde{u}_{n-1}^\eps)^{\perp_\eps}\subset
L^2_\eps$ by construction,  (\ref{v}) yields
\be\label{i2} \tilde{\mu}_n^\eps\leq
\mu_n+\mathcal{O}(\eps^{2/3-\delta}). \ee
>From (i)$_\text{k}$ and
(ii)$_\text{k}$, which are satisfied for $k\leq n-1$ thanks to the
recursion assumption, we have
\be\label{revne}
R^\eps(v_n^\eps)=\frac{Q^\eps(\chi_\eps u_n, \chi_\eps
u_n)-\underset{k=1}{\overset{n-1}{\sum}}\tilde{\mu}_k^\eps\left<\chi_\eps
u_n,
    \tilde{u}_k^\eps\right>_\eps^2}{\|\chi_\eps u_n\|_\eps^2-\underset{k=1}{\overset{n-1}{\sum}}\left<\chi_\eps u_n,
    \tilde{u}_k^\eps\right>_\eps^2}=\frac{Q^\eps(\chi_\eps u_n, \chi_\eps u_n)+\mathcal{O}(\eps^{2/3-\delta})}{\|\chi_\eps u_n\|_\eps^2+\mathcal{O}(\eps^{2/3-\delta})}.
\ee
Next,
\begin{eqnarray}\label{chieune}
\|\chi_\eps
u_n\|_\eps^2&=&\int_{-\frac{\eps^{-2/3}}{2}}^{\frac{\eps^{-2/3}}{2}}\frac{\chi_\eps^2u_n^2}{(1-\eps^{2/3}y)^{1/2}}dy\nonumber\\
&=&(1+\mathcal{O}(\eps^{2/3-\delta}))\int_{-\eps^{-\delta}}^{\eps^{-\delta}}u_n^2dy+\int_{\eps^{-\delta}\leq
  |y|\leq
  \frac{\eps^{-2/3}}{2}}\frac{\chi_\eps^2u_n^2}{(1-\eps^{2/3}y)^{1/2}}dy.
\end{eqnarray}
The last term in the right hand side of (\ref{chieune}) is estimated as follows
\begin{eqnarray}\label{chieune2}
\int_{\eps^{-\delta}\leq
  |y|\leq
  \frac{\eps^{-2/3}}{2}}\frac{\chi_\eps^2u_n^2}{(1-\eps^{2/3}y)^{1/2}}dy\leq
\sqrt{2}\int_{|y|\geq\eps^{-\delta}}u_n^2dy\lesssim \exp(-2\eps^{-\delta})\lesssim\eps^{2/3},
\end{eqnarray}
where we have used the following Lemma.
\begin{lem}\label{decayinf1}
For every $m\geq 1$, there exists a constant
$C_m>0$ such that for every $y\in \R$,
\be\label{decayinfun}
|u_m(y)|\leq C_m\exp(-|y|)
\ee
and
\be\label{decayinfun'}
|u_m'(y)|\leq C_m(|y|+1)\exp(-|y|).
\ee
\end{lem}

\begin{Proof}
Since $W_0(y)\to +\infty$ as
$y\to\infty$, we can fix $b_n>0$ such that $\inf\{W_0(y):|y|\geq
b_n\}>4+\mu_n$. Then,
$$(-4\partial_y^2+W_0(y)-\mu_n)e^{-|y|}=(W_0(y)-\mu_n-4)e^{-|y|}\geq
0,\quad |y|>b_n.$$
Since $u_n$ solves the eigenvalue problem
$$(-4\partial_y^2+W_0(y)-\mu_n)u_n=0,\quad y\in \R,$$
thanks to Corollary 2.8 in \cite{A}, there exists $C>0$ such that
$$|u_n(y)|\leq C e^{-|y|},\quad |y|\geq b_n+1.$$
Bound (\ref{decayinfun}) follows, since $u_n\in {\rm Dom}(M_0)\subset
H^1(\R)\subset L^\infty(\R)$. Then, from the differential equation
$M_0 u_n=\mu_n u_n$ and thanks to the asymptotic
behaviour of $W_0$, we infer
\be\label{un''}
|u_n''(y)|=\frac{1}{4}|\mu_n u_n(y)-W_0(y)u_n(y)|\lesssim
(|y|+1)e^{-|y|},\quad y\in\R.
\ee
By integration of (\ref{un''}) between $-\infty$ and $y$, we deduce, for $y<0$,
$$|u_n'(y)|=\left|\int_{-\infty}^y u_n''(s)ds\right|\lesssim (|y|+1)e^{-|y|}.$$
The same kind of estimate is obtained for $y>0$ by integration of
(\ref{un''}) between $y$ and $+\infty$, which provides
(\ref{decayinfun'}) and completes the proof of
Lemma \ref{decayinf1}.
\end{Proof}

Using Lemma \ref{decayinf1} again, as well as the normalization of
$u_n$, we infer that
\begin{eqnarray}\label{chieune3}
\int_{-\eps^{-\delta}}^{\eps^{-\delta}}u_n^2dy=1+\mathcal{O}(\eps^{2/3}).
\end{eqnarray}
>From (\ref{chieune}), (\ref{chieune2}) and (\ref{chieune3}), we
deduce that \be\label{denomrevne} \|\chi_\eps
u_n\|_\eps^2=1+\mathcal{O}(\eps^{2/3-\delta}). \ee On the other
side,
\begin{eqnarray}\label{qechieun}
Q^\eps(\chi_\eps u_n, \chi_\eps
u_n)&=&\int_{J_\eps}\left[4(1-\eps^{2/3}y)^{1/2}|\partial_y(\chi_\eps
  u_n)|^2+\frac{W_\eps|\chi_\eps u_n|^2}{(1-\eps^{2/3}y)^{1/2}}\right]dy\nonumber\\
&=& 4\int_{J_\eps}(1-\eps^{2/3}y)^{1/2}\chi_\eps'^2
  u_n^2dy+8\int_{J_\eps}(1-\eps^{2/3}y)^{1/2}\chi_\eps'\chi_\eps
  u_n'u_ndy\nonumber\\
&&+4\int_{-\eps^{-\delta}}^{\eps^{-\delta}}(1-\eps^{2/3}y)^{1/2}
  u_n'^2dy+4\int_{\eps^{-\delta}\leq |y|\leq
    \frac{\eps^{-2/3}}{2}}(1-\eps^{2/3}y)^{1/2}\chi_\eps^2  u_n'^2dy\nonumber\\
&&+\int_{-\eps^{-\delta/2}}^{\eps^{-\delta/2}}\frac{W_\eps u_n^2}{(1-\eps^{2/3}y)^{1/2}}dy+\int_{\eps^{-\delta/2}\leq |y|\leq
    \frac{\eps^{-2/3}}{2}}\frac{W_\eps|\chi_\eps u_n|^2}{(1-\eps^{2/3}y)^{1/2}}dy.
\end{eqnarray}
The first two integrals in the right hand side of
(\ref{qechieun}) are $\mathcal{O}(\eps^{2/3})$, because
$u_n\in H^1(\R)$,
$$\|\chi_\eps'\|_{L^\infty(J_\eps)}\lesssim \eps^{2/3}$$
and
$$\max\left\{(1-\eps^{2/3}y)^{1/2}:y\in {\rm Supp}\chi_\eps
\right\}\leq\sqrt{3/2}.$$
The fourth and last integrals in the right hand side of
(\ref{qechieun}) are also $\mathcal{O}(\eps^{2/3})$, thanks to Lemma
\ref{decayinf1}. From Lemma \ref{decayinf1}, we also infer that
\be\label{aa}
\int_{-\eps^{-\delta}}^{\eps^{-\delta}}(1-\eps^{2/3}y)^{1/2}u_n'^2dy=(1+\mathcal{O}(\eps^{2/3-\delta}))\int_{-\eps^{-\delta}}^{\eps^{-\delta}}
  u_n'^2dy=\int_{-\infty}^{+\infty}u_n'^2dy+\mathcal{O}(\eps^{2/3-\delta}).
\ee
>From Theorem \ref{theorem-main} and from the decay properties
of the function $\nu_n$ for $n\geq 1$ provided in (H$_\text{n}$),
we deduce that $\nu_\eps=\nu_0+\eps^{2/3}r_\eps$, where
$r_\eps=\mathcal{O}_{L^\infty(\R)}(1)$ and
$\nu_0r_\eps=\mathcal{O}_{L^\infty(\R)}(1)$ as $\eps\to 0$. As a
result, $W_\eps-W_0=3(\nu_\eps^2-\nu_0^2)\in L^\infty(\R)$, and
\be\label{wew0} \| W_\eps-W_0\|_{L^\infty(\R)}\lesssim \eps^{2/3}.
\ee Then, since $W_0(y)=\mathcal{O}(y)$ as $y\to \pm\infty$,
\be\label{bb}
\left\|\frac{W_\eps}{(1-\eps^{2/3}y)^{1/2}}-W_0\right\|_{L^\infty(-\eps^{-\delta/2},\eps^{-\delta/2})}\lesssim
\eps^{2/3-\delta}. \ee As a result, using once more Lemma
\ref{decayinf1}, \be\label{318bis}
\int_{-\eps^{-\delta/2}}^{\eps^{-\delta/2}}\frac{W_\eps
  u_n^2}{(1-\eps^{2/3}y)^{1/2}}dy=\int_{-\infty}^{+\infty}W_0
u_n^2dy+\mathcal{O}(\eps^{2/3-\delta}).
\ee
Finally, we get from (\ref{revne}), (\ref{denomrevne}),
(\ref{qechieun}), (\ref{aa}), (\ref{318bis}) and the estimates on the other
term in the right hand side of (\ref{qechieun}):
\be\label{revnef}
R^\eps(v_n^\eps)=R(u_n)+\mathcal{O}(\eps^{2/3-\delta})=\mu_n+\mathcal{O}(\eps^{2/3-\delta}),
\ee
which completes the proof of (\ref{v}) and of its corollary
(\ref{i2}).

\vspace{0.2cm}

{\bf Step 2. Asymptotic behaviour of the eigenfunction
  $\tilde{u}_n^\eps$.}
Property (i)$_\text{n}$ will be obtained as a consequence of
(\ref{i2}) and of the converse inequality
\be
\mu_n\leq \tilde{\mu}_n^\eps+\mathcal{O}(\eps^{2/3-\delta}).\nonumber
\ee
The proof of the latter inequality is delivered in Step 3 below.
The proof uses the following properties of the
eigenfunction $\tilde{u}_n^\eps$ corresponding to the $n^{\text{th}}$
eigenvalue $\tilde{\mu}_n^\eps$ of $\tilde{M}_\eps$.
\begin{lem}\label{decayinf2}
There exists a constant
$\tilde{C}_{n}>0$ such that for every $y\in J_\eps$ and $\eps>0$
sufficiently small,
\be\label{decayinfune}
|\tilde{u}_n^\eps(y)|\leq
\tilde{C}_{n}e^{-|y|},
\ee
whereas
\be\label{decayinfune'1}
|(\tilde{u}_n^\eps) '(y)|\leq\left\{
\begin{array}{lll}\tilde{C}_{n}(|y|+1)e^{-|y|}&\text{ if }& y\leq 0,\\
\tilde{C}_{n}(|y|+1)e^{-|y|}+\exp\left(-\frac{\eps^{-2/3}}{4}\right)&\text{ if }& 0<y\leq \frac{\eps^{-2/3}}{2},\\
  \frac{\tilde{C}_{n} \exp\left(-\frac{\eps^{-2/3}}{4}\right)}{(1-\eps^{2/3}y)^{1/2}}&\text{ if }& \frac{\eps^{-2/3}}{2}<y\leq
  \eps^{-2/3}.\end{array}\right.
\ee
\end{lem}

\begin{Proof}
In order to
prove (\ref{decayinfune}), we come back to the eigenfunction
$$\phi_{2n}^\eps(x)=\tilde{u}_n^\eps\left(\frac{1-x^2}{\eps^{2/3}}\right)$$
of $L_+^\eps$ corresponding to the eigenvalue
$\lambda_{2n}^\eps=\tilde{\mu}_n^\eps\eps^{2/3}$. Since
$$\|\tilde{u}_n^\eps\|_{H^1_\eps}^2=Q^\eps(\tilde{u}_n^\eps,\tilde{u}_n^\eps)=\tilde{\mu}_n^\eps\|\tilde{u}_n^\eps\|_\eps^2=\tilde{\mu}_n^\eps,$$
it follows from (\ref{i2}) and Lemma \ref{emb} that for $\eps$
sufficiently small,
\be\label{philinf}
\|\phi_{2n}^\eps\|_{L^\infty(\R)}=\|\tilde{u}_n^\eps\|_{L^\infty(J_\eps)}\leq
C\sqrt{\tilde{\mu}_n^\eps}\leq
C\sqrt{\mu_n+\mathcal{O}(\eps^{2/3-\delta})}\leq c_n,
\ee
where $c_n>0$ is an $\eps$-independent constant. Since $W_0(y)\gtrsim |y|$ as
$y\to \pm\infty$, we can fix $a_n$ large enough such that
$\inf\{W_0(y):|y|\geq a_n\}>4+\mu_n$. Then, using (\ref{wew0}) and
(\ref{i2}), we obtain, for
$x^2<1-a_n\eps^{2/3}$ and for $\eps$ small enough,
\be\label{c3}
\lefteqn{(-\eps^2\partial_x^2+x^2-1+3\eta_\eps^2-\eps^{2/3}\tilde{\mu}_n^\eps)\exp\left(-\frac{1-x^2}{\eps^{2/3}}\right)}\nonumber\\
&
= &
\eps^{2/3}\left(-2\eps^{2/3}-4x^2+W_\eps\left(\frac{1-x^2}{\eps^{2/3}}\right)-\tilde{\mu}_n^\eps\right)\exp\left(-\frac{1-x^2}{\eps^{2/3}}\right)\nonumber\\
&\geq
&\eps^{2/3}\left(-4+\inf\{W_0(y):y\geq
  a_n\}-\mu_n+\mathcal{O}(\eps^{2/3-\delta})\right)\exp\left(-\frac{1-x^2}{\eps^{2/3}}\right)\geq 0.
\ee
On the other side, $\phi_{2n}^\eps$
solves the differential equation
\be\label{odephi}
(-\eps^2\partial_x^2+x^2-1+3\eta_\eps^2-\eps^{2/3}\tilde{\mu}_n^\eps)\phi_{2n}^\eps=0.
\ee
Thus,
\be
(-\eps^2\partial_x^2+x^2-1+3\eta_\eps^2-\eps^{2/3}\tilde{\mu}_n^\eps)\psi_{n\pm}^\eps\geq
  0,\quad |x|<(1-a_n\eps^{2/3})^{1/2},
\ee
where
$$\psi_{n\pm}^\eps(x)=c_n
\exp\left(a_n-\frac{1-x^2}{\eps^{2/3}}\right)\pm\phi_{2n}^\eps(x).$$
Moreover, from (\ref{philinf}), we get
$$\psi_{n\pm}^\eps(\pm(1-a_n\eps^{2/3})^{1/2})\geq 0.$$
As a result, since for $\eps$ small enough, we also have like in (\ref{c3})
\be
x^2-1+3\eta_\eps^2-\eps^{2/3}\tilde{\mu}_n^\eps &=
&\eps^{2/3}\left(W_\eps\left(\frac{1-x^2}{\eps^{2/3}}\right)-\tilde{\mu}_n^\eps\right)\nonumber\\
&\geq &\eps^{2/3}\left(\inf\{W_0(y):y\geq
  a_n\}-\mu_n+\mathcal{O}(\eps^{2/3-\delta})\right)> 0,
\ee
the maximum principle ensures that
$$\psi_{n\pm}^\eps(x)\geq 0,\quad |x|<(1-a_n\eps^{2/3})^{1/2},$$
which is equivalent to
$$|\phi_{2n}^\eps(x)|\leq c_n
\exp\left(a_n-\frac{1-x^2}{\eps^{2/3}}\right), \quad |x|<(1-a_n\eps^{2/3})^{1/2}.$$
In terms of $\tilde{u}_n^\eps$, it means that
\be\label{inegun+inf}
|\tilde{u}_n^\eps(y)|\leq c_ne^{a_n}e^{-y}, \quad a_n\leq y\leq
\eps^{-2/3}.
\ee
On the other side, for $|x|\geq (1+a_n\eps^{2/3})^{1/2}$ and for $\eps$
sufficiently small, we obtain like in (\ref{c3})
\be
\lefteqn{(-\eps^2\partial_x^2+x^2-1+3\eta_\eps^2-\eps^{2/3}\tilde{\mu}_n^\eps)\exp\left(-\frac{x^2-1}{\eps^{2/3}}\right)}\nonumber\\
&
= &
\eps^{2/3}\left(2\eps^{2/3}-4x^2+W_\eps\left(\frac{1-x^2}{\eps^{2/3}}\right)-\tilde{\mu}_n^\eps\right)\exp\left(-\frac{x^2-1}{\eps^{2/3}}\right)\nonumber\\
&\geq &
\eps^{2/3}\left(-4+W_0(\frac{1-x^2}{\eps^{2/3}})\left(4\eps^{2/3}\frac{\frac{1-x^2}{\eps^{2/3}}}{W_0(\frac{1-x^2}{\eps^{2/3}})}+\frac{W_\eps\left(\frac{1-x^2}{\eps^{2/3}}\right)}{W_0(\frac{1-x^2}{\eps^{2/3}})}\right)-\mu_n+\mathcal{O}(\eps^{2/3})\right)\exp\left(-\frac{x^2-1}{\eps^{2/3}}\right)\nonumber\\
&\geq &0.
\ee
Thus, $\exp\left(-\frac{x^2-1}{\eps^{2/3}}\right)$ is a positive,
continuous supersolution of
$$(-\eps^2\partial_x^2+x^2-1+3\eta_\eps^2-\eps^{2/3}\tilde{\mu}_n^\eps)\phi=0$$
in $\{x:|x|>(1+a_n\eps^{2/3})^{1/2}\}$. From a slightly modified version of
Corollary 2.8 in \cite{A}, we deduce that
$$|\phi_{2n}^\eps(x)|\leq
2c_n\exp\left(1+a_n-\frac{x^2-1}{\eps^{2/3}}\right), \quad
|x|\geq (1+(a_n+1)\eps^{2/3})^{1/2}.$$
More precisely, the constant $2c_ne^{(a_n+1)}$ above has been chosen in such
a way that the inequality holds for $|x|= (1+(a_n+1)\eps^{2/3})^{1/2}$,
and the result in \cite{A} ensures that then, the inequality holds for
any $x$ such that $|x|\geq (1+(a_n+1)\eps^{2/3})^{1/2}.$
In terms of $\tilde{u}_n^\eps$, it means that
\be\label{inegun-inf}
|\tilde{u}_n^\eps(y)|\leq2c_ne^{a_n+1}e^y,\quad y\leq -(a_n+1).
\ee
Then, (\ref{decayinfune}) follows from (\ref{philinf}),
(\ref{inegun+inf}) and (\ref{inegun-inf}). We next prove
(\ref{decayinfune'1}). From (\ref{decayinfune}) and the differential
equation
$\tilde{M}^\eps\tilde{u}_n^\eps=\tilde{\mu}_n^\eps\tilde{u}_n^\eps$, we
infer that for every $y\in J_\eps$,
\be\label{d2une}
\left|\left(\partial_y(1-\eps^{2/3}y)^{1/2}\partial_y\tilde{u}_n^\eps\right)(y)\right|\leq\frac{\tilde{C}_ne^{-|y|}}{4(1-\eps^{2/3}y)^{1/2}}\left(\mu_n+\left(\underset{y\in\R}{\sup}\frac{W_0(y)}{|y|}\right)|y|+\mathcal{O}(\eps^{2/3-\delta})\right),\
\
\ee
where we have also used (\ref{i2}) and (\ref{wew0}). The estimate (\ref{decayinfune'1}) in the
case $y<0$ directly follows by integration of (\ref{d2une}) between $-\infty$ and $y$:
\be
\left|(\tilde{u}_n^\eps)'(y)\right|\leq\left|(1-\eps^{2/3}y)^{1/2}(\tilde{u}_n^\eps)'(y)\right|=\left|\int_{-\infty}^y\left(\partial_y(1-\eps^{2/3}y)^{1/2}\partial_y\tilde{u}_n^\eps\right)(s)ds\right|\lesssim
(|y|+1)e^{-|y|}.\ \ 
\ee
As for the case $0<y<\frac{\eps^{-2/3}}{2}$, integration of (\ref{d2une}) between
$y$ and $\frac{\eps^{-2/3}}{2}$ gives
\be\label{e2}
\left|(1-\eps^{2/3}y)^{1/2}(\tilde{u}_n^\eps)'(y)-\frac{1}{\sqrt{2}}
(\tilde{u}_n^\eps)'(\frac{\eps^{-2/3}}{2})\right|\lesssim (|y|+1)e^{-|y|},
\ee
which provides thanks to the triangular inequality
\be\label{u'ey}
\left|(\tilde{u}_n^\eps)'(y)\right|\lesssim
(|y|+1)e^{-|y|}+\left|(\tilde{u}_n^\eps)'(\frac{\eps^{-2/3}}{2})\right|.
\ee
Using basic integration, we also have
\be\label{e42}
\lefteqn{\tilde{u}_n^\eps(\frac{\eps^{-2/3}}{2})-\tilde{u}_n^\eps(\frac{\eps^{-2/3}}{4})}\nonumber\\
&=
&\!\!\!\!\!\!\int_{\frac{\eps^{-2/3}}{4}}^{\frac{\eps^{-2/3}}{2}}\!\left((\tilde{u}_n^\eps)'(s)-\frac{(\tilde{u}_n^\eps)'(\frac{\eps^{-2/3}}{2})}{\sqrt{2}(1-\eps^{2/3}s)^{1/2}}\right)ds+\frac{(\tilde{u}_n^\eps)'(\frac{\eps^{-2/3}}{2})}{\sqrt{2}}\int_{\frac{\eps^{-2/3}}{4}}^{\frac{\eps^{-2/3}}{2}}\!\!\frac{1}{(1-\eps^{2/3}s)^{1/2}}ds.\
\ \ \ \ \
\ee
Since the last integral in the right hand side of (\ref{e42}) is
bounded from below by $\frac{\eps^{-2/3}\sqrt{2}}{4}$, we deduce from
(\ref{e42}), (\ref{e2}) and (\ref{decayinfune}) that
\be\label{u'e4}
\left|(\tilde{u}_n^\eps)'(\frac{\eps^{-2/3}}{2})\right|\lesssim
\exp(-\frac{\eps^{-2/3}}{4}).
\ee
Combining (\ref{u'e4}) and (\ref{u'ey}), we get (\ref{decayinfune'1})
in the case when $0<y<\frac{\eps^{-2/3}}{2}$. Finally, we consider the case
when $\frac{\eps^{-2/3}}{2}<y<\eps^{-2/3}$. Integration of (\ref{d2une})
between $\frac{\eps^{-2/3}}{2}$ and $y$ yields
\be
\left|(\tilde{u}_n^\eps)'(y)\right| & \leq & \frac{1}{(1-\eps^{2/3}y)^{1/2}}\left(\sqrt{\frac{1}{2}}\left|(\tilde{u}_n^\eps)'(\frac{\eps^{-2/3}}{2})\right|+\int_{\frac{\eps^{-2/3}}{2}}^y
  \frac{(|s|+1)e^{-|s|}}{(1-\eps^{2/3}s)^{1/2}}ds\right)\nonumber\\
& \lesssim &
\frac{1}{(1-\eps^{2/3}y)^{1/2}}\left(\exp(-\frac{\eps^{-2/3}}{4})+\eps^{-2/3}\exp(-\frac{\eps^{-2/3}}{2})\int_{\frac{\eps^{-2/3}}{2}}^{\eps^{-2/3}}\frac{1}{(1-\eps^{2/3}s)^{1/2}}ds\right)\nonumber\\
& \lesssim & \frac{\exp(-\frac{\eps^{-2/3}}{4})}{(1-\eps^{2/3}y)^{1/2}},
\ee
where we have also used (\ref{u'e4}). This completes the proof of
(\ref{decayinfune'1}) and the proof of Lemma \ref{decayinf2}.
\end{Proof}

\vspace{0.2cm}

{\bf Step 3. Lower bound on $\tilde{\mu}_n^\eps$ and proof of (i)$_\text{n}$.}
In order to show that (i)$_\text{n}$ holds, we next prove the converse inequality
\be\label{i1}
\mu_n\leq \tilde{\mu}_n^\eps+\mathcal{O}(\eps^{2/3-\delta}),
\ee
which will be deduced from (\ref{ray}) and
\be\label{vt}
R\left(\tilde{v}_n^\eps\right)=\tilde{\mu}_n^\eps+\mathcal{O}(\eps^{2/3-\delta}), &\text{where}&
\tilde{v}_n^\eps=\chi_\eps \tilde{u}_n^\eps-\sum_{k=1}^{n-1}\left<\chi_\eps
    \tilde{u}_n^\eps,u_k\right>u_k,
\ee
In order to prove
(\ref{vt}), we proceed similarly as for the proof of (\ref{v}). First,
since (iii)$_\text{k}$ is assumed to be satisfied for $k\leq n-1$,
\be\label{revne2}
R(\tilde{v}_n^\eps) & = & \frac{Q(\chi_\eps \tilde{u}_n^\eps, \chi_\eps \tilde{u}_n^\eps)-\underset{k=1}{\overset{n-1}{\sum}}\mu_k\left<\chi_\eps \tilde{u}_n^\eps,
    u_k\right>^2}{\|\chi_\eps \tilde{u}_n^\eps\|^2-\underset{k=1}{\overset{n-1}{\sum}}\left<\chi_\eps \tilde{u}_n^\eps,
    u_k\right>^2}\nonumber\\
&= &\frac{Q(\chi_\eps \tilde{u}_n^\eps, \chi_\eps \tilde{u}_n^\eps)-\mu_{n-1}\left<\chi_\eps \tilde{u}_{n}^\eps,
    u_{n-1}\right>^2+\mathcal{O}(\eps^{2/3-\delta})}{\|\chi_\eps \tilde{u}_n^\eps\|^2-\left<\chi_\eps \tilde{u}_{n}^\eps,
    u_{n-1}\right>^2+\mathcal{O}(\eps^{2/3-\delta})}.
\ee
Then, thanks to Lemma \ref{decayinf2} and the normalization of $\tilde{u}_n^\eps$,
\begin{eqnarray}\label{chieune22}
\|\chi_\eps
\tilde{u}_n^\eps\|^2&=&\int_{-\frac{\eps^{-2/3}}{2}}^{\frac{\eps^{-2/3}}{2}}\chi_\eps^2|\tilde{u}_n^\eps|^2dy\nonumber\\
&=&(1+\mathcal{O}(\eps^{2/3-\delta}))\int_{-\eps^{-\delta}}^{\eps^{-\delta}}\frac{|\tilde{u}_n^\eps|^2}{(1-\eps^{2/3}y)^{1/2}}dy+\int_{\eps^{-\delta}\leq |y|\leq
  \frac{\eps^{-2/3}}{2}}\chi_\eps^2|\tilde{u}_n^\eps|^2dy.\nonumber\\
&=&(1+\mathcal{O}(\eps^{2/3-\delta}))\int_{J_\eps}\frac{|\tilde{u}_n^\eps|^2}{(1-\eps^{2/3}y)^{1/2}}dy+\mathcal{O}(\eps^{2/3})\nonumber\\
&=&1+\mathcal{O}(\eps^{2/3-\delta}).
\end{eqnarray}
Similarly, using Lemma \ref{decayinf2} and (\ref{bb}) and proceeding as in (\ref{qechieun}), we get
\begin{eqnarray}\label{qechieunt}
Q(\chi_\eps \tilde{u}_n^\eps, \chi_\eps
\tilde{u}_n^\eps)&=&\int_{-\infty}^{+\infty}\left(4|\partial_y(\chi_\eps
  \tilde{u}_n^\eps)|^2+W_0|\chi_\eps
  \tilde{u}_n^\eps|^2\right)dy\nonumber\\
&=& 4\int_{-\infty}^{+\infty}\chi_\eps'^2
  |\tilde{u}_n^\eps|^{2}dy+8\int_{-\infty}^{+\infty}\chi_\eps'\chi_\eps
  (\tilde{u}_n^\eps)'\tilde{u}_n^\eps dy\nonumber\\
&&+4(1+\mathcal{O}(\eps^{2/3-\delta}))\int_{-\eps^{-\delta}}^{\eps^{-\delta}}(1-\eps^{2/3}y)^{1/2}
  |(\tilde{u}_n^\eps)'|^2dy\nonumber\\
&&+4\int_{\eps^{-\delta}\leq |y|\leq
    \frac{\eps^{-2/3}}{2}}\chi_\eps^2  |(\tilde{u}_n^\eps)'|^2dy\nonumber\\
&&+\int_{-\eps^{-\delta/2}}^{\eps^{-\delta/2}}W_0 |\tilde{u}_n^\eps|^2dy+\int_{\eps^{-\delta/2}\leq |y|\leq
    \frac{\eps^{-2/3}}{2}}W_0|\chi_\eps \tilde{u}_n^\eps|^2dy\nonumber\\
&=& 4(1+\mathcal{O}(\eps^{2/3-\delta}))\int_{J_\eps}(1-\eps^{2/3}y)^{1/2}
  |(\tilde{u}_n^\eps)'|^2dy\nonumber\\
&&+\int_{-\eps^{-\delta/2}}^{\eps^{-\delta/2}}\frac{W_\eps|\tilde{u}_n^\eps|^2}{(1-\eps^{2/3}y)^{1/2}}dy+\mathcal{O}(\eps^{2/3-\delta})\nonumber\\
&=& Q^\eps(\tilde{u}_n^\eps,\tilde{u}_n^\eps)+\mathcal{O}(\eps^{2/3-\delta})=\tilde{\mu}_n^\eps+\mathcal{O}(\eps^{2/3-\delta}).
\end{eqnarray}
In order to deduce (\ref{vt}) from (\ref{revne2}), it remains to
estimate the scalar product $\left<\chi_\eps \tilde{u}_{n}^\eps,
    u_{n-1}\right>$. Notice that in the case when $n=1$, this term does
  not exists, and there is then nothing to do. From (iv)$_\text{n-1}$, there exists $c_{n-1}\in
  \R$ such that
\be\label{aprroxeigenv}
\|\chi_\eps \tilde{u}_{n-1}^\eps-c_{n-1}
  u_{n-1}\|\lesssim\eps^{1/3-\delta/2}.
\ee
Then, by triangular inequality, and thanks to (\ref{chieune22}) for
$n$ replaced by $n-1$,
\be\label{ti0}
||c_{n-1}|-1|\leq\|c_{n-1}
u_{n-1}-\chi_\eps\tilde{u}_{n-1}^\eps\|+\left|\|\chi_\eps\tilde{u}_{n-1}^\eps\|-1\right|\lesssim
\eps^{1/3-\delta/2},
\ee
whereas
\be\label{ti}
\lefteqn{|c_{n-1}|\left|\left<\chi_\eps \tilde{u}_{n}^\eps,
    u_{n-1}\right>\right|}\nonumber\\
&\leq &\left|\left<\chi_\eps
    \tilde{u}_{n}^\eps,c_{n-1} u_{n-1}-\chi_\eps
    \tilde{u}_{n-1}^\eps\right>\right|+\left|\left<\left(\chi_\eps^2-\frac{1}{(1-\eps^{2/3}y)^{1/2}}\right)
    \tilde{u}_{n}^\eps,\tilde{u}_{n-1}^\eps\right>\right|+\left|\left<\tilde{u}_{n}^\eps,\tilde{u}_{n-1}^\eps\right>_\eps\right|\nonumber\\
&\lesssim &\eps^{1/3-\delta/2},
\ee
where the first term in the right hand side of (\ref{ti}) has been
estimated thanks to the Cauchy-Schwarz inequality, (\ref{aprroxeigenv}) and (\ref{chieune22}). The second term has been estimated
thanks to Lemma \ref{decayinf2} for $\tilde{u}_{n}^\eps$ and for
$\tilde{u}_{n-1}^\eps$, and the last one is equal to 0. We deduce from
(\ref{ti0}) and (\ref{ti}) that
\be\label{ti2}
\left<\chi_\eps \tilde{u}_{n}^\eps,
    u_{n-1}\right>=\mathcal{O}(\eps^{1/3-\delta/2}).
\ee
Then, (\ref{vt}) and (\ref{i1}) follow from (\ref{revne2}), (\ref{chieune22}),
(\ref{qechieunt}) and (\ref{ti2}). Property (i)$_\text{n}$ is a direct consequence of (\ref{i1}) and (\ref{i2}).

\vspace{0.2cm}

{\bf Step 4. Proof of (ii)$_\text{n}$ and (iv)$_\text{n}$.} From
the definition of $\tilde{v}_n^\eps$ in (\ref{vt}), it is clear
that
$$\tilde{v}_n^\eps\in {\rm Span}(u_1,\cdots
,u_{n-1})^\perp.$$ Thus, $\tilde{v}_n^\eps$ can be decomposed as
\be\label{decvt} \tilde{v}_n^\eps=c_n^\eps u_n+w_n^\eps, \quad
\text{where} \quad c_n^\eps\in\R \text{ and }w_n^\eps\in {\rm
Span}(u_1,\cdots ,u_n)^\perp. \ee From (\ref{vt}) and
(i)$_\text{n}$, we have
$$\mu_n+\mathcal{O}(\eps^{2/3-\delta})=\tilde{\mu}_n^\eps+\mathcal{O}(\eps^{2/3-\delta})=R(\tilde{v}_n^\eps)=\frac{({c}_n^\eps)^2\mu_n
  +Q({w}_n^\eps,{w}_n^\eps)}{({c}_n^\eps)^2
  +\|{w}_n^\eps\|^2}\geq \frac{({c}_n^\eps)^2\mu_n
  +\|{w}_n^\eps\|^2\mu_{n+1}}{({c}_n^\eps)^2
  +\|{w}_n^\eps\|^2}.$$
It follows that
\be\label{prew}
(\mu_{n+1}-\mu_n)\|{w}_n^\eps\|^2\lesssim
\eps^{2/3-\delta}\|\tilde{v}_n^\eps\|^2.
\ee
Thanks to the definition of $\tilde{v}_n^\eps$ in (\ref{vt}),
property (iii)$_\text{k}$ for $k\leq n-1$ as well as (\ref{ti2}),
\be\label{tvctu}
\|\tilde{v}_{n}^\eps-\chi_\eps\tilde{u}_n^\eps\|\lesssim\eps^{1/3-\delta/2}.
\ee
On the other side, (\ref{chieune22}) ensures that $\|\chi_\eps \tilde{u}_n^\eps\|\to 1$
as $\eps \to 0$. As a result, $\|\tilde{v}_{n}^\eps\|\to 1$ as $\eps\to 0$, and
(\ref{prew}) implies
\be\label{ww}
\|w_n^\eps\|\lesssim\eps^{1/3-\delta/2}.
\ee
Moreover, from Lemmas \ref{decayinf1} and \ref{decayinf2}, we infer
that for any $k\geq 1$,
\be\label{scaprod}
\left<\chi_\eps\tilde{u}_n^\eps,u_k\right>&=&\int_{-\eps^{-\delta}}^{\eps^{-\delta}}\chi_\eps\tilde{u}_n^\eps
u_k dy+\int_{\eps^{-\delta}\leq|y|\leq\frac{\eps^{-2/3}}{2}}\chi_\eps\tilde{u}_n^\eps
u_k dy\nonumber\\
&=&(1+\mathcal{O}(\eps^{2/3-\delta}))\int_{J_\eps}\frac{\chi_\eps
  u_k\tilde{u}_n^\eps}{(1-\eps^{2/3}y)^{1/2}}dy+\mathcal{O}(\eps^{2/3}).
\ee
>From (\ref{ww}) and (\ref{scaprod}) we deduce in particular
that for every $k\geq n+1$, \be\label{scpd}
\mathcal{O}(\eps^{2/3})+\left<\chi_\eps
  u_k,\tilde{u}_n^\eps\right>_\eps(1+\mathcal{O}(\eps^{2/3-\delta}))=\left<\chi_\eps\tilde{u}_n^\eps,u_k\right>=\left<\tilde{v}_n^\eps,u_k\right>=\left<w_n^\eps,u_k\right>=\mathcal{O}(\eps^{1/3-\delta/2}),
\ee
which proves (ii)$_\text{n}$. Then, (iv)$_\text{n}$ is a consequence of
the triangular inequality, (\ref{tvctu}) and (\ref{ww}):
$$\|\chi_\eps\tilde{u}_n^\eps-c_n^\eps
u_n\|\leq\|\chi_\eps\tilde{u}_n^\eps-\tilde{v}_n^\eps\|+\|w_n^\eps\|\lesssim \eps^{1/3-\delta/2}.
$$

\vspace{0.2cm}

{\bf Step 5. Proof of (iii)$_\text{n}$ and (v)$_\text{n}$.}
Like in (\ref{decvt}), we decompose $v_{n-1}^\eps$ as
$$
v_{n-1}^\eps=\tilde{c}_{n-1}^\eps \tilde{u}_{n-1}^\eps+\tilde{w}_{n-1}^\eps, \quad
\text{where} \quad \tilde{c}_{n-1}^\eps\in\R \text{ and
}\tilde{w}_{n-1}^\eps\in {\rm Span}(\tilde{u}_1^\eps,\cdots
,\tilde{u}_{n-1}^\eps)^{\perp_\eps}.
$$
>From (\ref{v}) for $n$
replaced by $n-1$ and (i)$_\text{n-1}$, we have \be
\tilde{\mu}_{n-1}^\eps+\mathcal{O}(\eps^{2/3-\delta})=\mu_{n-1}+\mathcal{O}(\eps^{2/3-\delta})=R^\eps({v}_{n-1}^\eps)
& =&\frac{(\tilde{c}_{n-1}^\eps)^2\tilde{\mu}_{n-1}^\eps
  +Q^\eps(\tilde{w}_{n-1}^\eps,\tilde{w}_{n-1}^\eps)}{(\tilde{c}_{n-1}^\eps)^2
  +\|\tilde{w}_{n-1}^\eps\|_\eps^2}\nonumber\\
&\geq &\frac{(\tilde{c}_{n-1}^\eps)^2\tilde{\mu}^\eps_{n-1}
  +\|\tilde{w}_{n-1}^\eps\|_\eps^2\tilde{\mu}_{n}^\eps}{(\tilde{c}_{n-1}^\eps)^2
  +\|\tilde{w}_{n-1}^\eps\|_\eps^2}.\nonumber
\ee
Using (i)$_\text{n}$ and (i)$_\text{n-1}$, it follows that
\be\label{prewt}
(\mu_{n}-{\mu}_{n-1}+\mathcal{O}(\eps^{2/3-\delta}))\|\tilde{w}_{n-1}^\eps\|_\eps^2=(\tilde{\mu}^\eps_{n}-\tilde{\mu}_{n-1}^\eps)\|\tilde{w}_{n-1}^\eps\|_\eps^2\lesssim
\eps^{2/3-\delta}\|{v}_{n-1}^\eps\|_\eps^2.
\ee
Thanks to the definition of $v_{n-1}^\eps$ given by (\ref{v}) and
property (ii)$_\text{k}$ for $k\leq n-2$,
\be\label{vcu}
\|v_{n-1}^\eps-\chi_\eps u_{n-1}\|_\eps\lesssim\eps^{1/3-\delta/2}.
\ee
Thanks to (\ref{denomrevne}) for $n$ replaced by $n-1$, $\|\chi_\eps{u}_{n-1}\|_\eps\to
1$ as $\eps \to 0,$ thus $\|v_{n-1}^\eps\|_\eps\to 1$ as $\eps\to
0$. As a result, we deduce from (\ref{prewt}) that
\be\label{tw}
\|\tilde{w}_{n-1}^\eps\|_\eps\lesssim\eps^{1/3-\delta/2}.
\ee
Then, for every $k\geq n$, we get
\be\label{355}
\left<\chi_\eps\tilde{u}_k^\eps,{u}_{n-1}\right>(1+\mathcal{O}(\eps^{2/3-\delta}))=\left<\chi_\eps
  u_{n-1},\tilde{u}_k^\eps\right>_\eps=\left<{v}_{n-1}^\eps,\tilde{u}_k^\eps\right>_\eps=\left<\tilde{w}_{n-1}^\eps,\tilde{u}_k^\eps\right>_\eps=\mathcal{O}(\eps^{1/3-\delta/2}),
\ee
using similar arguments as in the derivation of
(\ref{scpd}). Moreover,
\be\label{356}
\left<\chi_\eps u_{n-1},\tilde{u}_k^\eps\right>_\eps
&=&(1+\mathcal{O}(\eps^{2/3-\delta}))\left(\left<\chi_\eps\tilde{u}_k^\eps,
    u_{n-1}\right>-\int_{\eps^{-\delta}\leq|y|\leq\frac{\eps^{-2/3}}{2}}\chi_\eps
    u_{n-1}\tilde{u}_k^\eps dy\right)\nonumber\\
&&+\int_{\eps^{-\delta}\leq|y|\leq\frac{\eps^{-2/3}}{2}}\frac{\chi_\eps
  u_{n-1}\tilde{u}_k^\eps}{(1-\eps^{2/3}y)^{1/2}}dy\nonumber\\
&=&(1+\mathcal{O}(\eps^{2/3-\delta}))\left<\chi_\eps\tilde{u}_k^\eps,
    u_{n-1}\right>+\mathcal{O}(\eps^{2/3}),
\ee
where the two integrals in the right hand side of (\ref{356}) have
been estimated thanks to the Cauchy-Schwarz inequality, Lemma
\ref{decayinf1} and the normalization condition
$\|\tilde{u}_k^\eps\|_\eps=1$. The combination of (\ref{355}) and (\ref{356})
completes the proof of (iii)$_\text{n}$. Then, (v)$_\text{n}$ follows from
the triangular inequality, (\ref{vcu}) and (\ref{tw}):
$$\|\chi_\eps u_{n-1}-\tilde{c}_{n-1}^\eps
\tilde{u}_{n-1}^\eps\|_\eps\leq\|\chi_\eps u_{n-1}-v_{n-1}^\eps\|_\eps+\|\tilde{w}_{n-1}^\eps\|_\eps\lesssim \eps^{1/3-\delta/2}.
$$
It completes the proof of (G$_\text{n}$), and therefore the proof of
Theorem \ref{theorem-eigenvalues}.
\end{Proof}
\section{Semi-classical limit for eigenvalues of $L^{\eps}_+$}

We list here formal results of the semi-classical theory that describe 
the distribution of eigenvalues of $L_+^{\eps}$. We will show that the
standard Bohr--Sommerfeld quantization rule does not give the
correct asymptotic behavior of the eigenvalues of $L_+^{\eps}$ as $\eps \to 0$
because the potential $V_{\eps}(x)$ depends on $\eps$.
Nevertheless, the Bohr--Sommerfeld quantization rule gives the
correct scaling ${\cal O}(\eps^{2/3})$ in agreement with the asymptotic limit 
(\ref{asymptotic-limit-eigenvalues}) in Theorem \ref{theorem-eigenvalues}.

Eigenvalue problem for operator $L^{\eps}_+$ can be rewritten in
the form
\begin{equation}
\label{double-well-problem}
\left(-\partial_x^2 + \eps^{-2} V_{\eps}(x) \right) u(x) = \eps^{-2} \lambda  u(x), \quad x \in \R.
\end{equation}
By properties of $\eta_{\eps}$ following from Theorem
\ref{theorem-main}, the potential $V_{\eps}(x)$ has the properties
\begin{itemize}
\item $V_{\eps}(x) \in {\cal C}^{\infty}(\mathbb{R})$ for any small $\eps > 0$,

\item $\underset{\eps \to 0}{\lim} V_{\eps}(x) =V_{0}(x) $, where $V_0\in {\cal
    C}(\mathbb{R})$ is given by
$$
V_0(x)= \left\{ \begin{array}{lr} 2(1-x^2), \quad & |x| \leq 1, \\
x^2 - 1, \quad & |x| \geq 1, \end{array} \right.
$$

\item $V_{\eps}(x)$ takes its absolute minimum at $\pm
a_{\eps}$ for any small $\eps \geq 0$ and $a_\eps\to 1$ as $\eps \to 0$,

\item $V_{\eps}(x) \to +\infty$ as $|x| \to \infty$ for any small $\eps \geq 0$.
\end{itemize}

If $V_{\eps}(x)$ is replaced by $V_0(x)$, the eigenvalue problem
(\ref{double-well-problem}) takes a simplified form 
\begin{equation}
\label{double-well-problem-simplified}
\left(-\partial_x^2 + \eps^{-2} V_0(x) \right) u(x) = \eps^{-2} \lambda  u(x), \quad x \in \R,
\end{equation}
which describes the eigenvalues of the operator $\tilde{L}_+^{\eps}$
mentioned in section 1. As it is well-known 
(see a recent review in \cite{Dobr}), the eigenvalues of the
Schr\"odinger operator $-\partial_x^2 + \eps^{-2} V(x)$, with a
smooth, $\eps$-independent double well potential $V(x)$, are twice
degenerate in the semi-classical limit $\eps \to 0$. Namely, the
eigenvalues are grouped by pairs. In each pair, the two eigenvalues
are exponentially close one from another as $\eps\to 0$. The
asymptotic distribution of these pairs of eigenvalues is determined by the
Bohr--Sommerfeld quantization rule.  

Let us try to apply the Bohr-Sommerfeld quantization rule to the eigenvalue problems 
(\ref{double-well-problem}) and (\ref{double-well-problem-simplified})
for the operators $L_+^\eps$ and $\tilde{L}_+^\eps$, in spite of
the fact that this rule was proved rigorously 
by Fedoryuk \cite{Fedoryuk} only for a class of $\eps$-independent, analytic
potentials. Since neither (\ref{double-well-problem}) nor (\ref{double-well-problem-simplified}) 
satisfies assumptions of the main theorem in \cite{Fedoryuk}, this application is purely 
formal. According to the standard Bohr--Sommerfeld rule, the consequent eigenvalues 
$\lambda_{2n-1}^{\eps}$ and $\lambda_{2n}^{\eps}$ of the Schr\"{o}dinger equation 
(\ref{double-well-problem}) with the double-well potential $V_{\eps}(x)$ 
would be given asymptotically by
\begin{equation}
\label{BSrule}
\int_{x_-^{\eps}(\lambda)}^{x_+^{\eps}(\lambda)} \sqrt{ \lambda -
V_{\eps}(x)} dx \sim \eps \pi \left( n - \frac{1}{2} \right), \quad
\mbox{\rm as} \quad \eps \to 0, \;\; \mbox{\rm for fixed} \;\; n \geq 1,
\end{equation}
where $x_{\pm}^{\eps}(\lambda)$ are the roots of $V_{\eps}(x) = \lambda$ on
$\R_+$, such that $0 < x_-^{\eps}(\lambda) <1< x_+^{\eps}(\lambda) < \infty$. Let us use the scaling
\begin{equation}
\label{BSrule-scaling}
y = \frac{1-x^2}{\eps^{2/3}}, \quad V_{\eps}(x) = \eps^{2/3}
W_{\eps}(y), \quad \lambda = \eps^{2/3} \mu,
\end{equation}
where $W_{\eps}(y) = 3 \nu_{\eps}^2(y) - y$ and $\mu$ is a new
eigenvalue. The Bohr--Sommerfeld rule is rewritten in an equivalent form by
$$
\int_{y_-^{\eps}(\mu)}^{y_+^{\eps}(\mu)} \frac{\sqrt{ \mu -
W_{\eps}(y)}}{\sqrt{1 - \eps^{2/3} y}} dy \sim \pi (2n -1), \quad
\mbox{\rm as} \quad \eps \to 0, \;\; \mbox{\rm for fixed} \;\; n \geq 1,
$$
where $y_{\pm}^{\eps}(\mu)$ are the roots of $W_{\eps}(y) = \mu$ on $\R$, such that 
$-\infty < y_-^{\eps}(\mu) < 0 < y_+^{\eps}(\mu) < \infty$. Taking the limit 
$\eps \to 0$ for a fixed $n \geq 1$, we obtain 
\begin{equation}
\label{BSrule-scaled}
\int_{y_-(\mu)}^{y_+(\mu)} \sqrt{ \mu - W_0(y)} dy \sim \pi (2n -1), \quad
\mbox{\rm for fixed} \;\; n \geq 1,
\end{equation}
where $W_0(y) = 3 \nu_0^2(y) - y$ and $y_{\pm}(\mu)$ are the roots of $W_0(y) = \mu$ on $\R$.
The new expression is the Bohr--Sommerfeld quantization rule for
the Schr\"{o}dinger operator $M_0 = -4 \partial_y^2 + W_0$ 
and it is only valid for large $n \gg 1$. Therefore, the 
Bohr--Sommerfeld quantization rule (\ref{BSrule}) 
does not recover the statement of Theorem \ref{theorem-eigenvalues} correctly. 
Meantime, it still implies that the eigenvalues $\lambda_{2n-1}^{\eps}$ and $\lambda_{2n}^{\eps}$ for
a fixed $n \geq 1$ are scaled as ${\cal O}(\eps^{2/3})$ as $\eps \to 0$. 
The discrepancy of the Bohr--Sommerfeld rule is explained by the fact that 
the smooth potential $V_{\eps}(x)$ in the eigenvalue problem (\ref{double-well-problem}) 
depends on $\eps$. 

Note that the limit $\eps \to 0$ can be computed exactly for the
simplified eigenvalue problem (\ref{double-well-problem-simplified}) 
thanks to the scaling transformation (\ref{BSrule-scaling}). In this case, 
the limiting formula (\ref{BSrule-scaled}) holds with 
$W_0(y)$ replaced by $2y$ for $y \geq 0$ and $-y$ for $y \leq 0$, so that 
$y_-(\mu) = -\mu$ and $y_+(\mu) = \mu/2$. In other words,
$$
\int_{-\mu}^{0} \sqrt{ \mu + y} dy + \int_{0}^{\mu/2} 
\sqrt{ \mu - 2y} dy \sim \pi (2n -1), \quad
\mbox{\rm for fixed} \;\; n \geq 1,
$$
and the computations of integrals gives $\mu_n \sim \left( \pi (2n-1) \right)^{2/3}$,
in agreement with the behavior ${\cal O}(n^{2/3})$ of eigenvalues
of the Schr\"{o}dinger operator with a linearly growing potential as $|y| \to \infty$ \cite{Sukumar}.
Therefore, the Bohr-Sommerfeld quantization rule suggests that 
the eigenvalues $\{ \tilde{\lambda}_n^{\eps} \}_{n \geq 1}$ 
of the simplified operator $\tilde{L}_+^{\eps}$ considered in our 
previous work \cite{GalPel} satisfy the asymptotic limit 
\begin{equation}
\label{asymptotic-limit-eigenvalues-BSrule}
\lim_{\eps \downarrow 0} \frac{\tilde{\lambda}_{2n-1}^\eps}{\eps^{2/3}} = 
\lim_{\eps \downarrow 0} \frac{\tilde{\lambda}_{2n}^\eps}{\eps^{2/3}} = 
\left( \pi (2n-1) \right)^{2/3}, \quad
\mbox{\rm for fixed} \;\; n \geq 1.
\end{equation}
However, the justification of the asymptotic limit (\ref{asymptotic-limit-eigenvalues-BSrule}) 
cannot rely on the work of Fedoryuk \cite{Fedoryuk} because the 
$\eps$-independent potential $V_0(x)$ in the simplified 
eigenvalue problem (\ref{double-well-problem-simplified})
is continuous but not $\mathcal{C}^1$ on $\R$.  

\section{Proof of Lemma \ref{asyop}}

Let $\alpha > 1$ be like in the assumption of the lemma, and $A =
\|x^\alpha f\|_{L^\infty(\R_+)} < \infty$. We first prove
(\ref{asympphi}) by contradiction. We proceed as follows. We
suppose that (\ref{asympphi}) is not true. Namely, we make the assumption\vspace{.2cm}\\
(G$_\alpha$)\hfill $\phi(x)\neq\mathcal{O}(x^{-(\alpha+1)})$, \hfill
$x^\alpha f\in L^\infty(A_+,+\infty)$\hfill\null\vspace{.2cm}\\
If $\alpha>2$, we prove that (G$_\alpha$) implies (G$_{\alpha-2}$),
such that after a finite number of steps, (G$_\alpha$) implies
(G$_{\tilde{\alpha}}$) for some $\tilde{\alpha}\in (0,2]$. On the other
side, we show that for $0<\alpha\leq 2$, (G$_\alpha$) yields to a
contradiction.

If (\ref{asympphi}) is not true,
then, up to a change of $f$ and $\phi$ into $-f$ and $-\phi$, there
exists a sequence $(x_n)_{n\geq n_0}$ (where $n_0>A$), such that $x_n\uparrow\infty$,
$x_n\geq A_+$ and
$$
x_n^\alpha W(x_n)\phi(x_n)>n.
$$
Then,
$$x_n^\alpha \phi''(x_n)=x_n^\alpha W(x_n)\phi(x_n)-x_n^\alpha
f(x)\geq x_n^\alpha W(x_n)\phi(x_n)-A>n-A.$$
For $n\geq n_0>A$, we define
$$y_n=\sup\{y>x_n, \forall x\in (x_n,y), x^\alpha
W(x)\phi(x)-A>(n-A)/2\}.$$
By continuity of $W$ and $\phi$, for every $n\geq n_0$, either
$y_n=+\infty$ or
\be\label{phiyn}
\phi(y_n) & =& \frac{n+A}{2y_n^\alpha W(y_n)}.
\ee
We distinguish the two following cases:\\
\null\hfill\begin{tabular}{ll}
A)& There exists $n_1\geq n_0$ such that $y_{n_1}=+\infty$\\
B)& For every $n\geq n_0$, $y_n<+\infty$.
\end{tabular}\hfill\null\\
In case B), extracting a subsequence of $(x_n)_{n\geq n_0}$ if
necessary, one can assume that
$$x_{n_0}<y_{n_0}<x_{n_0+1}<y_{n_0+1}<x_{n_0+2}<\cdots$$
For $n\geq n_0+1$, we define
$$\tilde{x}_n=\inf\{y<x_n,\forall x\in (y,x_n), x^\alpha
W(x)\phi(x)-A>3(n-A)/4\}.$$
Since $y_{n-1}<x_n$ and
$$y_{n-1}^\alpha
W(y_{n-1})\phi(y_{n-1})-A=(n-1-A)/2<3(n-A)/4,$$ we deduce
$\tilde{x}_n>y_{n-1}>-\infty$. Moreover, by continuity,
$\phi(\tilde{x}_n)=(3n+A)/(4\tilde{x}_n^\alpha W(\tilde{x}_n)),$
and $\phi(x)>
(3n+A)/(4x^\alpha W(x))$ for $x>\tilde{x}_n$, $x$ close to
$\tilde{x}_n$. Therefore
\begin{eqnarray*}
\phi'(\tilde{x}_n)\geq\frac{3n+A}{4}\frac{d}{dx}\left(\frac{1}{x^\alpha
    W(x)}\right)_{|x=\tilde{x}_n}&\geq
&-\frac{3n+A}{4}\left(\alpha+\frac{\|W'\|_{L^\infty}}{C_+}\right)\frac{1}{\tilde{x}_n^{\alpha+1}W(\tilde{x}_n)}\\
&\geq &
-C_1 n\frac{1}{\tilde{x}_n^{\alpha+1}W(\tilde{x}_n)},
\end{eqnarray*}
for some $C_1>0$. By definition of $y_n$ and $\tilde{x}_n$, for every $x\in (\tilde{x}_n,y_n)$,
$$x^\alpha\phi''(x)\geq\frac{n-A}{2}.$$
Thus,
\begin{eqnarray}\label{phiprime}
\phi'(x) & \geq &
\phi'(\tilde{x}_n)+\frac{n-A}{2}\int_{\tilde{x}_n}^x\frac{1}{y^\alpha}dy\nonumber\\
&\geq &-C_1
n\frac{1}{\tilde{x}_n^{\alpha+1}W(\tilde{x}_n)}+\frac{n-A}{2}\int_{\tilde{x}_n}^x\frac{1}{y^\alpha}dy=:G_n(x).
\end{eqnarray}
Notice that $G_n(\tilde{x}_n)<0$, whereas
$$G_n(+\infty)=\ \left\{\begin{array}{ll}+\infty& \text{ if }\alpha\leq
    1\\  g_n&\text{ if }\alpha>   1,\end{array}\right.$$
where for $\alpha>1$,
$$g_n\sim \frac{n-A}{2(\alpha-1)\tilde{x}_n^{\alpha-1}}>0\text{ as }
n\to +\infty.$$
As a result, for $n$ sufficiently large, since $G_n$ is increasing on
$(\tilde{x}_n,+\infty)$, $G_n$ vanishes exactly once on that
interval. Moreover, this unique zero $z_n$ of $G_n$ is defined by
$$ \int_{\tilde{x}_n}^{z_n}\frac{1}{y^\alpha}dy=\frac{2C_1 n}{n-A}\frac{1}{\tilde{x}_n^{\alpha+1}W(\tilde{x}_n)},$$
thus
$$z_n=\tilde{x}_n+\mathcal{O}\left(\frac{1}{\tilde{x}_n^2}\right).$$
By integration of (\ref{phiprime}), we infer that for $x\in (\tilde{x}_n,y_n)$,
\begin{eqnarray*}
\phi(x) &\geq & \phi(\tilde{x}_n)+\int_{\tilde{x}_n}^x G_n(y)dy\\
&\geq & \phi(\tilde{x}_n)+\int_{\tilde{x}_n}^{z_n} G_n(y)dy\\
&\geq & \phi(\tilde{x}_n)-\frac{C_1
  n}{\tilde{x}_n^{\alpha+1}W(\tilde{x}_n)}(z_n-\tilde{x}_n)\\
&\geq &
\frac{3n+A}{4}\frac{1}{\tilde{x}_n^{\alpha}W(\tilde{x}_n)}-\frac{C_2n}{\tilde{x}_n^{\alpha+3}W(\tilde{x}_n)},
\end{eqnarray*}
for some constant $C_2>0$. Therefore, for $n$ large
enough, for every $x\in (\tilde{x}_n,y_n)$, since $W$ is increasing on $(A_+,+\infty)$,
$$\phi(x)\geq
\frac{5n}{8}\frac{1}{\tilde{x}_n^{\alpha}W(\tilde{x}_n)}\geq\frac{5n}{8}\frac{1}{x^{\alpha}W(x)}.$$
For $n$ sufficiently large, $5n/8>(n+A)/2$, and it provides a
contradiction with (\ref{phiyn}), which means that case B) can not happen. In case A), for every $x\geq x_{n_1}$,
\be\label{phi''below}
x^\alpha\phi''(x)\geq (n_1-A)/2>0.
\ee
Therefore $\phi'(x)\uparrow 0$ as
$x\uparrow \infty$, otherwise $\phi$ would not be in $L^2(\R)$. Thus,
for every $x\geq x_{n_1}$, $\phi'(x)\leq 0$, and therefore
$\phi(x)\downarrow 0$ as $x\uparrow \infty$. If $0<\alpha\leq 1$,
(\ref{phi''below}) provides a contradiction with the fact that
$\phi'(x)\to 0$ as $x\to\infty$. If $\alpha>1$, integration of
(\ref{phi''below}) between $x$ and $+\infty$ yields
\be\label{phi'below}
-\phi'(x)\geq\frac{n_1-A}{2(\alpha-1)}x^{1-\alpha}.
\ee
This is a contradiction with $\phi(x)\to 0$, if $1<\alpha\leq
2$. Finally, if $\alpha>2,$ by integration of (\ref{phi'below}),
$$\phi(x)\geq\frac{n_1-A}{2(\alpha-1)}\int_x^{+\infty}y^{1-\alpha}dy
=\frac{n_1-A}{2(\alpha-1)(\alpha-2)}x^{2-\alpha}.$$
Thus,
$$\phi(x)\underset{x\to\infty}{\neq}\mathcal{O}\left(x^{-(\alpha-2)-1}\right).$$
Since the assumption $x^\alpha f\in L^\infty(A_+,+\infty)$ implies
$x^{\alpha-2}f\in L^\infty(A_+,+\infty)$, we have proved that (G$_\alpha$) implies
(G$_{\tilde{\alpha}}$) if $\alpha>2$. The proof of (\ref{asympphi}) is
completed by induction. Then, since $\phi''=W\phi-f$, we deduce
\be\label{asympphi''}
\phi''(x) = \mathcal{O}(x^{-\alpha}). 
\ee
We next prove that 
\be
& \phantom{t} & \phi'(x) = \left\{\begin{array}{ll}\mathcal{O}(x^{-(\alpha-1)})& \text{ if }\alpha>1,\\
 o(1) &\text{ if }0<\alpha\leq 1.\end{array}\right.\label{asympphi'} 
\ee
By integration of (\ref{asympphi''}), if
$\alpha>1$, $\phi'(x)$ has a limit as $x\to +\infty$. This limit can
only be 0, because $\phi\in L^2$. (\ref{asympphi'}) is then obtained
by integration of (\ref{asympphi''}) between $x$ an $+\infty$. If
$\alpha\leq 1$, (\ref{asympphi'}) is a consequence of the fact that $\phi(x)\to 0$ and
$\phi''(x)\to 0$ as $x\to +\infty$.

Let $\chi\in \mathcal{C}^\infty(\R)$ be such
that
$$\chi(x)=\left\{\begin{array}{ll}0 & \text{ if } x\leq 1\\ 1 & \text{
      if } x\geq 2\end{array}\right.$$ For $m\in \N$, let $\phi_m, f_m\in
\mathcal{C}^\infty(\R)$ be the functions defined by
$$\phi_m(x)=\chi(x)x^{-(\alpha+\gamma m+1)}$$
and
$$f_m(x)=-\phi_m''(x)+W(x)\phi_m(x).$$
>From now on, we assume that $f$ and $W$ have asymptotic series
(\ref{serf}) as $x\to +\infty$, so that
\begin{eqnarray*}
f_m(x)&\underset{x\to +\infty}{\approx} &x^{-(\alpha+\gamma m)}\sum_{k=0}^{+\infty}v_kx^{-\gamma
    k}+(\alpha+\gamma m +1)(\alpha+\gamma m +2)x^{-(\alpha+\gamma (m
    +3/\gamma))}.\\
& \underset{x\to +\infty}{\approx} &x^{-(\alpha+\gamma m)}\sum_{k=0}^{+\infty}\tilde{v}_kx^{-\gamma
    k},
\end{eqnarray*}
where $\tilde{v}_k=v_k$ if $k\neq 3/\gamma$ and $\tilde{v}_{3/\gamma}=v_{3/\gamma}+(\alpha+\gamma m +1)(\alpha+\gamma m +2)$.
Notice also that the assumption $W(x)\geq C_+x$ implies $v_0\geq
C_+>0$. As a result, there exists coefficients $(\tilde{c}_m)_{m\in \N}$ such
that for every $M\geq 0$,
$$f(x)=\sum_{m=0}^M \tilde{c}_m f_m(x)+g_M(x),$$
where $g_M(x)=\mathcal{O}(x^{-\alpha-\gamma(M+1)})$ as $x\to
+\infty$. Then,
$$\phi(x)=(-\partial_x^2+W)^{-1}f(x)=\sum_{m=0}^M \tilde{c}_m
\phi_m(x)+\psi_M(x),$$
where $\psi_M=(-\partial_x^2+W)^{-1}g_M.$ Thanks to (\ref{asympphi}),
(\ref{asympphi'}) and (\ref{asympphi''}), for $M$ large enough,
$\psi_M(x)=\mathcal{O}(x^{-\alpha-\gamma(M+1)-1})$,
$\psi_M'(x)=\mathcal{O}(x^{-\alpha-\gamma(M+1)+1})$ and
$\psi_M''(x)=\mathcal{O}(x^{-\alpha-\gamma(M+1)})$. Since this is true
for arbitrarily large values of $M$, then (\ref{serphi}) and
(\ref{asympphi'2}) follow.

\section{Proof of Lemma \ref{V0below}}

By Proposition \ref{proposition-Painleve}, we know that
$\nu_0$ is a strictly increasing function on $\R$, with asymptotics at
$\pm\infty$ given by (\ref{asympnu0-}) and
(\ref{asympnu0+}). Moreover, $\nu_0$ has a unique inflection
point. From the behaviour of $\nu_0(y)$ as $y\to\pm\infty$, we infer
that $W_0(y)=3\nu_0(y)^2-y\to +\infty$ as $y\to \pm\infty$. We
are going to prove that the global minimum of $W_0$ is actually
strictly positive. We argue by contradiction. If it is not the case,
we can define
$$
y_1=\inf\{y>0,\ \nu_0(y)=\sqrt{y/3}\},
$$
where we recall that $W_0(y)>0$ if $y\leq 0$. By continuity,
$\nu_0(y_1)=\sqrt{y_1/3}$. We also denote the unique inflection
point of $\nu_0$ by $y_0$. Since $\nu_0>0$ solves (\ref{P2-eq}),
$y_0>0$ is the unique solution of the equation
$\nu_0(y_0)=\sqrt{y_0}$, and $\nu_0''(y)>0$ if $y<y_0$, whereas
$\nu_0''(y)<0$ if $y>y_0$. Notice that since $\nu_0(0)>0$ and
$\nu_0(y_1)=\sqrt{y_1/3}<\sqrt{y_1}$, we have necessarily
$0<y_0<y_1$. Moreover, since $\nu_0$ is strictly increasing, we
have $\sqrt{y_0}=\nu_0(y_0) < \nu_0(y_1)=\sqrt{y_1/3}$, and
therefore $0<3y_0<y_1.$

\vspace{0.2cm}

{\bf First step:  upper bound on $\mathbf{y_1}$.} For $y>0$, we introduce
the function $z(y)=\nu_0(y)/\sqrt{y}$ and rewrite (\ref{P2-eq}) in terms of $z(y)$ as
$$
z''(y)+\frac{1}{y}z'(y)
=\frac{yz(y)}{4}\left(z(y)^2-1+\frac{1}{y^3}\right).$$
Since $z(y)\to +\infty$ as $y\to 0^+$ and $z(y)\to 1$ as $y\to
+\infty$ with $z(y)<1$ for $y$ large enough (because for $y>y_0$,
$\nu_0''(y)<0$ and therefore
$\nu_0(y) < \sqrt{y}$), we deduce that $z(y)$
admits a global minimum at $y = y_m > 0$, where
$$0\leq z''(y_m)=\frac{y_m
  z(y_m)}{4}\left(z(y_m)^2-1+\frac{1}{y_m^3}\right).$$
The assumption of non-positivity of $W_0$ implies that $z(y_m)\leq
1/\sqrt{3}$. Thus,
$$
\frac{1}{y_m^3}\geq 1-z(y_m)^2\geq \frac{2}{3}.
$$
As a result, since $\nu_0(y_m)\leq\sqrt{y_m/3}$,
\be\label{uby1}
3y_0<y_1\leq y_m\leq\left(\frac{3}{2}\right)^{1/3}.
\ee

\vspace{0.2cm}

{\bf Second step:  upper bound on $\mathbf{\nu_0'(y_0)}$.}
Since $\nu_0$ is increasing on $\R$ and $\nu_0(y)^2-y>0$ if $y<y_0$,
we deduce, for every $y<y_0$,
\be
\nu_0'(y_0)-\nu_0'(y) & = &
\int_y^{y_0}\frac{\nu_0(t)}{4}\left(\nu_0(t)^2-t\right)dt\nonumber\\
&\leq &
\int_y^{y_0}\frac{\nu_0(y_0)}{4}\left(\nu_0(y_0)^2-t\right)dt=\frac{\sqrt{y_0}}{8}(y_0-y)^2.
\ee
By integration, it follows that for $y<y_0$,
\be\label{nua}
\nu_0(y) & = &\sqrt{y_0}-\int_y^{y_0}\nu_0'(t)dt\nonumber\\
&\leq &\sqrt{y_0}-\nu_0'(y_0)(y_0-y)+\frac{\sqrt{y_0}}{24}(y_0-y)^3.
\ee
The right hand side reaches its minimum (for $y<y_0$) at $y=y_p$, where $y_p<y_0$ is
defined by $(y_0-y_p)^2=8\nu_0'(y_0)/\sqrt{y_0}$, and (\ref{nua}) at
$y=y_p$ yields
$$\nu_0(y_p)\leq
\sqrt{y_0}-\frac{4\sqrt{2}}{3}\frac{\nu_0'(y_0)^{3/2}}{y_0^{1/4}}.$$
Since $\nu_0>0$, the right hand side has to be strictly
positive. Therefore
\be\label{ubnpy0}
\nu_0'(y_0)&\leq & \left(\frac{9}{32}\right)^{1/3}\sqrt{y_0}.
\ee

\vspace{0.2cm}

{\bf Third step: upper bound on $\mathbf{\nu_0'(y_1)}$.}
On the one side, notice that for $y>y_0$, $\nu_0''(y)<0$, and
therefore $\nu_0'(y_1)\leq\nu_0'(y_0).$ On the other side, if $y<y_1$,
$\nu_0(y)^2>y/3$, and $\nu_0(y_1)^2=y_1/3$, thus
$$
\nu_0'(y_1)\leq
\left.\frac{d}{dy}\sqrt{\frac{y}{3}}\right|_{y=y_1}=\frac{1}{2\sqrt{3 y_1}}.
$$
As a result, thanks to (\ref{ubnpy0}) and (\ref{uby1})
\be\label{ubnpy1}
\nu_0'(y_1)&\leq &\min\left(
  \left(\frac{9}{32}\right)^{1/3}\frac{\sqrt{y_1}}{\sqrt{3}},\frac{1}{2\sqrt{3 y_1}}\right).
\ee

\vspace{0.2cm}

{\bf Fourth step: upper bound on $\mathbf{\nu_0'(y)}$ for
$\mathbf{y>y_1}$.} For $\delta\in (0,2/3)$ to be fixed later, we
define
$$y_2(\delta)=\sup\{y>y_1, \forall t\in (y_1,y),\nu_0(t)^2\leq
(1-\delta)t\}$$
(notice that $\nu_0(y_1)^2=y_1/3<(1-\delta)y_1$). Then, for every
$y\in (y_1,y_2(\delta))$,
\be\label{ubnp}
\nu_0'(y) & = &
\nu_0'(y_1)+\int_{y_1}^y\frac{\nu_0(t)}{4}(\nu_0(t)^2-t)dt\nonumber\\
 & \leq &
 \nu_0'(y_1)+\int_{y_1}^y\frac{\nu_0(y_1)}{4}(-\delta t)dt\nonumber\\
& \leq & \nu_0'(y_1)-\frac{\nu_0(y_1)}{8}\delta(y^2-y_1^2).
\ee

\vspace{0.2cm}

{\bf Fifth step: bound from below on $\mathbf{y_2(\delta)}$.}
For $\delta\in (0,2/3)$, we introduce the function $h_\delta$ defined
for $y>y_1$ by
$$h_\delta(y):=\nu_0'(y_1)-\frac{\nu_0(y_1)}{8}\delta(y^2-y_1^2)-\frac{\sqrt{1-\delta}}{2\sqrt{y}}.$$
>From (\ref{ubnpy1}) and since $\delta<2/3$, we infer
$h_\delta(y_1)<0$. Thus, if we define
$$y_3(\delta):=\sup\{y>y_1,\forall t\in (y_1,y), h_\delta(t)<0\},$$
we deduce from (\ref{ubnp}) that for $y\in (y_1,\min(y_2(\delta),y_3(\delta)))$,
\be
\nu_0(y) -\sqrt{1-\delta}\sqrt{y}& = &\nu_0(y_1)
-\sqrt{1-\delta}\sqrt{y_1}+\int_{y_1}^yh_\delta(t)dt<0,
\ee
which implies that
\be\label{bby2}
y_3(\delta) & \leq  & y_2(\delta).
\ee

\vspace{0.2cm}

{\bf Sixth step: $\mathbf{y_3=+\infty}$.}
We shall see next
that for an appropriate choice of $\delta$, $y_3(\delta)=+\infty$,
which implies that $y_2(\delta)=+\infty$ thanks to (\ref{bby2}) . This
provides a contradiction with the assumption of non positivity of
$W_0$, since $\nu_0(y)\sim \sqrt{y}$ as $y\to+\infty$. An elementary
calculation shows that $h_\delta$ reaches its maximum (for $y>y_1$) at
$$y=y_M:=\left(\frac{\sqrt{3}\sqrt{1-\delta}}{\sqrt{y_1}\delta}\right)^{2/5}>y_1,$$
where the inequality comes from (\ref{uby1}) and from the fact that $\delta<2/3$. From (\ref{ubnpy1}), we
obtain
\be\label{ubhM}
h_\delta(y_M)\leq \min\left(
  \left(\frac{9}{32}\right)^{1/3}\frac{\sqrt{y_1}}{\sqrt{3}},\frac{1}{2\sqrt{3 y_1}}\right)+\frac{\delta
  y_1^{5/2}}{8\sqrt{3}}-\frac{5y_1^{1/10}\delta^{1/5}(1-\delta)^{2/5}}{8\cdot  3^{1/10}}.
\ee
For $\delta=1/3$, elementary calculations show that the right hand
side in (\ref{ubhM}) is strictly negative for any $y_1\in
(0,(3/2)^{1/3})$, which implies that
$y_3(1/3)=+\infty$ and completes the proof of the lemma.

\section{Proof of Lemma \ref{shen}}
We denote
$$U_\eps(z)=W_0(\eps^{-2/3}-\eps^{2/3}|z|^2),\quad z\in \R^d.$$
We are going to show that there exists a constant $C>0$ such that for $\eps>0$
sufficiently small, for every ball $B\subset \R^d$,
\be\label{estshen}
\underset{z\in B}{\max}\  U_\eps(z)\leq \frac{C}{|B|}\int_B U_\eps(z)dz.\ee
According to Theorem 0.3 in \cite{S}, Lemma \ref{shen} follows. First, we notice that,
thanks to Lemma \ref{V0below} and (\ref{W0inf}), there exist
$C_1,C_2>0$ such that for every $y\in \R$,
\be\label{C1C2}
C_1(1+|y|)\leq W_0(y)\leq C_2(1+|y|).
\ee
Given $z_0\in \R^d$ and $r>0$, as $z$ describes $B(z_0,r)$, $|z|$ describes the interval
$[|z_0|-r,|z_0|+r]$ if $|z_0|\geq r$ and the interval $[0,|z_0|+r]$ if
$|z_0|\leq r$. Since the function
$$f(s)=|\eps^{-2/3}-\eps^{2/3}s^2|,\quad s\in \R_+$$
is decreasing on $[0,\eps^{-2/3}]$ and increasing on
$[\eps^{-2/3},+\infty)$, we infer that $\max\{f(|z|), z\in B(z_0,r)\}$
can only take the three different values depending on $z_0$ and $r$: either
$$\underset{z\in
  B(z_0,r)}{\max}f(|z|)=\eps^{2/3}(|z_0|+r)^2-\eps^{-2/3}\quad
\text{and}\quad |z_0|+r\geq\eps^{-2/3}\quad \text{(case 1)},$$
or
$$\underset{z\in
  B(z_0,r)}{\max}f(|z|)=\eps^{-2/3}-\eps^{2/3}(|z_0|-r)^2\quad
\text{and}\quad 0\leq|z_0|-r\leq\eps^{-2/3}\quad\text{(case 2)},$$
or 
$$\underset{z\in  B(z_0,r)}{\max}f(|z|)=\eps^{-2/3}\quad
\text{and}\quad |z_0|-r\leq 0\quad \text{(case 3)}.$$
We are
next going to prove (\ref{estshen}) in each of these 3 cases.\\
{\bf Case 1.} We first show that for every $z_0$, $r$ like in case 1,
we have
\be\label{rz01}
|z_0|+\frac{ r}{\sqrt{2}}\geq \eps^{-2/3}.
\ee
Under the extra assumption
\be\label{exa}
\eps^{2/3}(|z_0|+r)^2-\eps^{-2/3}\geq\eps^{-2/3},
\ee
(\ref{rz01}) clearly holds. On the other side, if (\ref{exa}) is not
true, then $|z_0|-r>0$ since otherwise, $0\in
[|z_0|-r, |z_0|+r]$ and $\max\{f(|z|), z\in
B(z_0,r)\}\geq f(0)=\eps^{-2/3}$, contradicting the assumption that we
are in case 1. Then, we also have
$$\eps^{2/3}(|z_0|+r)^2-\eps^{-2/3}=f(|z_0|+r)\geq f(|z_0|-r)\geq \eps^{-2/3}-\eps^{2/3}(|z_0|-r)^2,$$
which can be rewritten as
$$|z_0|^2+r^2\geq \eps^{-4/3}.$$
Since $r<|z_0|$, we deduce
$$\eps^{-4/3}\leq |z_0|^2+r^2\leq |z_0|^2+\frac{r^2}{2}+\sqrt{2}r|z_0|=\left(|z_0|+\frac{r}{\sqrt{2}}\right)^2,$$
which means that (\ref{rz01}) also holds if (\ref{exa}) is not true. Let
$\alpha= \sqrt{3}/2\geq 1/2+\sqrt{2}/4.$ Then,
\begin{eqnarray}\label{case1}
\lefteqn{\left(\eps^{2/3}(|z_0|+\alpha
  r)^2-\eps^{-2/3}\right)-\frac{1}{2}\left(\eps^{2/3}(|z_0|+r)^2-\eps^{-2/3}\right)-\frac{1}{2}\left(\eps^{2/3}(|z_0|+\frac{r}{\sqrt{2}})^2-\eps^{-2/3}\right)}\nonumber\\
&=&2\eps^{2/3}|z_0|r(\alpha-\frac{1}{2}-\frac{1}{2\sqrt{2}})+\eps^{2/3}r^2(\alpha^2-1/2-1/4).\
\ \ \ \ \ \ \ \ \ \ \ \ \ \ \ \ \ \ \ \ \ \ \ \ \ \ \ \ \ \ \ \ \ \ \ 
\end{eqnarray}
We deduce from (\ref{rz01}) and (\ref{case1}) that for every $z\in B(z_0,r)$ such that
$|z|>|z_0|+\alpha r$,
\begin{eqnarray}\label{precase1}
f(|z|)&=&\eps^{2/3}|z|^2-\eps^{-2/3}\geq \eps^{2/3}(|z_0|+\alpha
r)^2-\eps^{-2/3}\nonumber\\
&\geq&
\frac{1}{2}\left(\eps^{2/3}(|z_0|+r)^2-\eps^{-2/3}\right) =\frac{1}{2}\underset{z\in
B(z_0,r)}{\max}f(|z|).
\end{eqnarray}
Then, we conclude thanks to (\ref{C1C2}) and (\ref{precase1}) that
\be\label{cas1}
\frac{1}{|B(z_0,r)|}\int_{B(z_0,r)}U_\eps(z)dz & \geq &
\frac{C_1}{|\mathbb{B}^d|r^d}\int_{B(z_0,r)}(1+f(|z|))dz\nonumber\\
& \geq &
\frac{C_1}{|\mathbb{B}^d|r^d}\int_{B(z_0,r)\string\ B(0,|z_0|+\alpha r)}(1+f(|z|))dz\nonumber\\
& \geq &
\frac{C_1}{|\mathbb{B}^d|r^d}\int_{\left\{z\in B(z_0,r):\
  z\cdot\frac{z_0}{|z_0|}\geq |z_0|+\alpha r)\right\}}(1+f(|z|))dz\nonumber\\
& \geq &
\frac{C_1v_\alpha}{2|\mathbb{B}^d|}\underset{z\in B(z_0,r)}{\max}(1+f(|z|))\nonumber\\
& \geq &
\frac{C_1v_\alpha}{2|\mathbb{B}^d|C_2}\ \underset{z\in B(z_0,r)}{\max}U_\eps(z),
\ee
where $v_\alpha$ denotes the volume of $\{z\in B(0,1): z_1\geq \alpha\}$.\\
{\bf Case 2.} 
The assumption that we are in case 2 implies
$$\eps^{-2/3}-\eps^{2/3}(|z_0|-r)^2=f(|z_0|-r)\geq f(|z_0|+r)\geq\eps^{2/3}(|z_0|+r)^2-\eps^{-2/3},$$
and thus
$$|z_0|^2\leq |z_0|^2+r^2\leq \eps^{-4/3}.$$
It follows that 
\begin{eqnarray*}
\left(\eps^{-2/3}-\eps^{2/3}(|z_0|-
  r/2)^2\right)-\frac{1}{2}\left(\eps^{-2/3}-\eps^{2/3}(|z_0|-r)^2\right)
&=&\frac{1}{2}\left(\eps^{-2/3}-\eps^{2/3}|z_0|^2\right)+\frac{\eps^{2/3}r^2}{4}\geq
0.
\end{eqnarray*}
We deduce that for every $z\in B(z_0,r)$ such that $|z|\leq
|z_0|-r/2$,
\be\label{case2}
f(|z|)\geq \frac{1}{2}f(|z_0|-r).
\ee
Then, we show that this last estimates holds as soon as $z\in
B(z_0,r)$ and $z\cdot z_0/|z_0|\leq |z_0|-7r/8$. Indeed, under this
assumption, Pythagoras' theorem ensures that
\be
|z|^2& = &\left(z\cdot\frac{z_0}{|z_0|}\right)^2+|z-z_0|^2-\left((z-z_0)\cdot\frac{z_0}{|z_0|}\right)^2\nonumber\\
&\leq
&\left(|z_0|-\frac{7r}{8}\right)^2+r^2-\left(\frac{7r}{8}\right)^2\nonumber\\
&=&\left(|z_0|-\frac{r}{2}\right)^2-\frac{3r}{4}(|z_0|-r)\leq\left(|z_0|-\frac{r}{2}\right)^2.\nonumber
\ee
Then, we conclude similarly as in case 1, thanks to (\ref{C1C2}) and (\ref{case2})
\be\label{cas2}
\frac{1}{|B(z_0,r)|}\int_{B(z_0,r)}U_\eps(z)dz & \geq &
\frac{C_1}{|\mathbb{B}^d|r^d}\int_{B(z_0,r)}(1+f(|z|))dz\nonumber\\
& \geq &
\frac{C_1}{|\mathbb{B}^d|r^d}\int_{\left\{z\in B(z_0,r):\ z\cdot\frac{z_0}{|z_0|}\leq |z_0|- 7r/8\right\}}(1+f(|z|))dz\nonumber\\
& \geq &
\frac{C_1v_{7/8}}{2|\mathbb{B}^d|}\underset{z\in B(z_0,r)}{\max}(1+f(|z|))\nonumber\\
& \geq &
\frac{C_1v_{7/8}}{2|\mathbb{B}^d|C_2}\ \underset{z\in B(z_0,r)}{\max}U_\eps(z).
\ee
{\bf Case 3.} First, we notice that the assumption that we are in case
3 yields
$$\eps^{-2/3}\geq f(|z_0|+r)\geq \eps^{2/3}(|z_0|+r)^2-\eps^{-2/3},$$
which gives
\be\label{z0+r}
|z_0|+r<\sqrt{2}\eps^{-2/3}.
\ee
Thus, since $|z_0|\leq r$, we get
\be\label{z03}
|z_0|\leq\frac{1}{\sqrt{2}}\eps^{-2/3}.
\ee
If the extra assumption 
\be\label{exa3}
r\geq 5|z_0|/4
\ee
holds, then (\ref{z03}) and the triangular inequality give $B(0,r/5)\subset B(z_0,r)$. Moreover, if $z\in
B(0,r/5)$, then we get from (\ref{z0+r})
\be\label{bz0}
\eps^{-2/3}-\eps^{2/3}|z|^2\geq 23\eps^{-2/3}/25.
\ee
Then, we conclude similarly as in cases 1 and 2:
\be\label{cas31}
\frac{1}{|B(z_0,r)|}\int_{B(z_0,r)}U_\eps(z)dz & \geq
&\frac{C_1}{|\mathbb{B}^d|r^d}\int_{B(0,r/5)}\left(1+23\eps^{-2/3}/25\right)dz\nonumber\\
& \geq
&\frac{23C_1}{25\cdot 3^d C_2}\ \underset{z\in
  B(z_0,r)}{\max}U_\eps(z).
\ee
As for the last case when (\ref{exa3}) is not true, we have then
$$\left\{z\in B(z_0,r): z\cdot \frac{z_0}{|z_0|}\leq\frac{7r}{40}\right\}\subset
B(z_0,r)\cap B(0,|z_0|).$$
Indeed, using also $|z_0|\leq r\leq\frac{5|z_0|}{4}$, we have then
\be
|z|^2
&=&\left(z\cdot\frac{z_0}{|z_0|}\right)^2+|z-z_0|^2-\left(|z_0|-z\cdot\frac{z_0}{|z_0|}\right)^2=|z-z_0|^2-|z_0|^2+2|z_0|z\cdot \frac{z_0}{|z_0|}\nonumber\\
&\leq
&r^2-|z_0|^2+\frac{7r|z_0|}{20}\leq |z_0|^2.
\ee
On the other side, for $z\in B(0,|z_0|),$ thanks to (\ref{z03}), we have 
$$f(|z|)=\eps^{-2/3}-\eps^{2/3}|z|^2\geq \eps^{-2/3}/2.$$
Then, we conclude similarly as in the previous cases:
\be\label{cas32}
\frac{1}{|B(z_0,r)|}\int_{B(z_0,r)}U_\eps(z)dz & \geq &
\frac{C_1}{|\mathbb{B}^d|r^d}\int_{B(z_0,r)}(1+f(|z|))dz\nonumber\\
& \geq &
\frac{C_1}{|\mathbb{B}^d|r^d}\int_{\left\{z\in B(z_0,r), z\cdot\frac{z_0}{|z_0|}\leq 7r/40\right\}}(1+f(|z|))dz\nonumber\\
& \geq &
\frac{C_1v_{33/40}}{2|\mathbb{B}^d|C_2}\ \underset{z\in B(z_0,r)}{\max}U_\eps(z).
\ee
>From (\ref{cas1}), (\ref{cas2}), (\ref{cas31}) and (\ref{cas32}), we
infer that (\ref{estshen}) holds, with
$$C=\min\left(\frac{C_1v_\alpha}{2|\mathbb{B}^d|C_2},\frac{C_1v_{7/8}}{2|\mathbb{B}^d|C_2},\frac{23C_1}{25\cdot
    3^d C_2},\frac{C_1v_{33/40}}{2|\mathbb{B}^d|C_2}\right),$$
which completes the Proof of the lemma.

\end{document}